\documentclass[aps,prd,reprint,twocolumn,groupedaddress]{revtex4}
\usepackage{amsmath,amsfonts,amssymb,bm}
\usepackage{graphicx}
\usepackage{color}
\usepackage{subfigure}
\usepackage{multirow}
\usepackage{textcomp}
\usepackage{slashed}
\usepackage{makecell}
\usepackage{enumitem}

\definecolor{purple}{rgb}{0.5,0,0.5}
\definecolor{blue}{rgb}{0.0,0,0.9}

\usepackage[colorlinks=true, pdfstartview=FitV, linkcolor=purple, citecolor= purple, urlcolor=blue]{hyperref}

\begin{document}


\title{Two-photon Transition Form Factor of $\eta_{c,b}(nS)$ and $\chi_{c0,b0}(nP)$ via Relativized Mock Meson States}




\author{Tian-Cheng Ding, Jian Huang, Muyang Chen}
\email{muyang@hunnu.edu.cn}
\affiliation{Department of Physics, Hunan Normal University, and Key Laboratory of Low-Dimensional Quantum Structures and Quantum Control of Ministry of Education, Changsha 410081, China}

\date{\today}

\begin{abstract}
We construct relativized mock meson states for heavy quarkonium, where the Dirac spinors import kinetic relativistic correction and the dynamic wave functions are the same as those solved from the Schrödinger equation. We find that the relativistic correction imported by Dirac spinors is crucial to study the two-photon transition form factors. Using relativized mock meson states, we give credible predictions for the two-photon transition form factors and decay widths of $\eta_{c,b}(n^1S_0)$ and $\chi_{c,b}(n^3P_0)$ ($n=1,2,3$).
\end{abstract}


\maketitle


\renewcommand{\thesection}{\Roman{section}}
\renewcommand{\thesubsection}{\Roman{section}-\arabic{subsection}}

\section{Introduction}\label{sec:introduction}

The two-photon transition form factor of heavy quarkonium system provides a powerful probe into the electromagnetic structure and internal dynamics of these bound states. These transitions, where a quarkonium state decays into two photons, are governed by Quantum Chromodynamics (QCD) and offer a unique opportunity to study strong interaction. The form factor describing the momentum dependence of the transition amplitude is sensitive to the spatial distribution of the quark-antiquark pair and relativistic corrections. Charmonium and bottomonium systems are particularly well-suited for such studies due to their relatively simple structure compared to light quark mesons.

Experimentally, two-photon transitions of charmonium have been extensively studied in CLEO \cite{Ecklund2008}, BABAR \cite{Lees2010}, BESIII \cite{Ablikim2012} and Belle \cite{Masuda2018}. These experiments have provided high precision measurements of decay widths $\Gamma_{\gamma\gamma}$ and the $Q^2$ dependence of the form factors. For example, the two-photon decay width of the $\eta_c$ is measured to be $ 5.1\pm 0.4 $ keV, the form factor is measured upto $Q^2 \approx 50$ GeV \cite{Lees2010}. While for $\eta_b$, two-photon decay is not seen due to significantly smaller decay width. Higher-precision study is expected in the future electron-ion colliders, see Ref. \cite{Bertulani2025}.

Theoretically, the two-photon transition form factor has been investigated using a variety of approaches, such as lattice QCD \cite{Dudek2006,Colquhoun2023,Meng2023}, Dyson-Schwinger equation approach \cite{Chen2017}, the sum rule \cite{Lucha2012}, non-relativistic QCD \cite{Feng2015,Sang2016,Yang2021}, perturbative QCD \cite{Cao1999,Hoferichter2020}, the light front approach \cite{Ryu2018,Babiarz2019,Li2022} and so on. In potential quark models, the quarkonium wave function is derived by solving the Schrödinger equation with a confining potential, however, the role of relativistic corrections and the precise shape of the quarkonium distribution amplitude are still areas of active research.

In this paper, we focus on the effect of the relativized mock meson state on the two-photon transition form factor of $\eta_{c,b}(nS)$ and $\chi_{c0,b0}(nP)$. This paper is organized as following. In section \ref{sec:model} we introduce the model used in this work. In section \ref{sec:resultsCC} we present and comment our results for the charmonium spectrum, wave functions and the two-photon transition form factors. Section \ref{sec:resultsBB} is parallel to section \ref{sec:resultsCC}, but for the bottomonium. Summary and conclusion is given in section \ref{sec:conclusion}. There are also two appendixes, \ref{sec:appendixA} formulates the derivation of the non-relativistic limit of the relativized mock meson state, and \ref{sec:appendixB} formulates the expressions of the two-photon transition form factors of the mesons.

\section{Quark Model and Mock Meson State}\label{sec:model}

In this paper we adopt almost the same quark model as Ref. \cite{Sun2023,Lakhina2006} with a little modifications. In the following we first give a brief description of the Halmitonian we used, then we will focus on the two kinds of representations of the meson and the expressions of two-photon transition form factor of pseudoscalar and scalar mesons.

The masses and wave functions are obtained by solving the radial Schrödinger equation,
\begin{equation}\label{eq:SchrödingerEq}
 (T + V - E)\varphi(r) = 0,
\end{equation}
where $T$ is the kinetic energy operator, $V$ is the potential between the quarks, $E$ is the energy of this system and $\varphi(r)$ is the radial wavefunction, $r$ is the distance between the quark and antiquark. The relativistic kinetic energy is
\stepcounter{equation}
\begin{equation}\tag{\theequation a}\label{eq:RelKEnergy}
 T =\sqrt{m^2 +\bm{k}^2}+\sqrt{\bar{m}^2 +\bar{\bm{k}}^2},
\end{equation}
where $m$ and $\bar{m}$ are the masses of the quark and antiquark, $\bm{k}$ and $\bar{\bm{k}}$ are the 3-momentum of the quark and antiquark. Note that in this paper a bold character stands for a 3-dimensional vector, for example, $\bm{k} = \vec{k}$. In the non-relativistic limit, the kinetic energy reduces into
\begin{equation}\tag{\theequation b}\label{eq:nRelKEnergy}
 T =m + \bar{m} + \frac{\bm{k}^2}{2m} + \frac{\bar{\bm{k}}^2}{2\bar{m}}.
\end{equation}

The potential could be decomposed into
\begin{equation}\label{eq:interaction}
 V = H^{\text{SI}} + H^{\text{SS}} + H^{\text{T}} + H^{\text{SO}}.
\end{equation}
$H^{\text{SI}}$ is the spin independent part, which is composed of a coulombic potential and a linear potential,
\begin{equation}\label{eq:interactionSI}
 H^{\text{SI}} = -\frac{4\alpha_s(Q^2)}{3r} + br,
\end{equation}
where $b$ is a constant and $\alpha_s(Q^2)$ is the running coupling of the strong interaction. The other three terms are spin dependent.
\begin{equation}\label{eq:interactionSS}
 H^{\text{SS}} = \frac{32\pi\alpha_s(Q^2)}{9m\bar{m}}\tilde{\delta}_\sigma(\bm{r}) \bm{s}\cdot\bm{\bar{s}}
\end{equation}
is the spin-spin contact hyperfine potential, where $\bm{s}$ and $\bm{\bar{s}}$ are the spin of the quark and antiquark respectively, and $\tilde{\delta}_\sigma(\bm{r}) = (\frac{\sigma}{\sqrt{\pi}})^3 \text{e}^{-\sigma^2 r^2}$ with $\sigma$ being a parameter.
\begin{equation}\label{eq:interactionT}
 H^{\text{T}} = \frac{4\alpha_s(Q^2)}{3m\bar{m}} \frac{1}{r^3}\left( 3\frac{(\bm{s}\cdot\bm{r})(\bm{\bar{s}}\cdot\bm{r})}{r^2} - \bm{s}\cdot\bm{\bar{s}} \right)
\end{equation}
is the tensor potential. $H^{\text{SO}}$ is the spin-orbit interaction potential
\begin{equation}
\label{eq:interactionSO}
 H^{\text{SO}}  =  \frac{\bm{S}\cdot\bm{L}}{2}\left[ (\frac{1}{2m^2} + \frac{1}{2\bar{m}^2})(\frac{4\alpha_s(Q^2)}{3r^3} - \frac{b}{r})+ \frac{8\alpha_s(Q^2)}{3m\bar{m}r^3} \right]
\end{equation}
where $\bm{S} = \bm{s} + \bm{\bar{s}}$, and $\bm{L}$ is the orbital angular momentum of the quark and antiquark system.

In equations (\ref{eq:interactionSI})$\sim$(\ref{eq:interactionSO}), the running coupling takes the following form \cite{Godfrey1985},
\begin{equation}\label{eq:runningAlphas}
 \alpha_s(Q^2) = \frac{4\pi}{\beta \log(\text{e}^{\frac{4\pi}{\beta\alpha_0}} + \frac{Q^2}{\Lambda^2_{\text{QCD}}})},
\end{equation}
where $\Lambda_{\text{QCD}}$ is the energy scale below which nonperturbative effects take over, $\beta = 11-\frac{2}{3}N_f$ with $N_f$ being the flavor number, $Q$ is the momentum transfer, and $\alpha_0$ is a constant. Eq. (\ref{eq:runningAlphas}) approaches the one loop running form of QCD at large $Q^2$ and saturates at low $Q^2$.

The potentials containing $\frac{1}{r^3}$, Eq. (\ref{eq:interactionT}) and Eq. (\ref{eq:interactionSO}) are divergent. A cutoff $r_c$ is introduced to eliminate the divergence, i.e. $\frac{1}{r^3} \to \frac{1}{r^3_c}$ for $r \leq r_c$. Herein $r_c$ is a parameter to be fixed by observables.

$N_f$ and $\Lambda_{\text{QCD}}$ are chosen according to QCD estimation.  In principle,  $\Lambda_\text{QCD}$ depends on the choice of $N_f$. However, we found that $\Lambda_{\text{QCD}}$ varying in $[0.2,0.4]\text{ GeV}$ induces small changes in the physical quantities ($\lesssim 4\%$) \cite{Sun2023}. In this work we fix $\Lambda_{\text{QCD}} = 0.3 \text{ GeV}$, $N_f=4$ for charmonium and $N_f = 5$ for bottomium. $m_c$ and $m_b$ are the same as Ref. \cite{Sun2023}. So there are 4 free parameters in all: $\alpha_0$, $b$, $\sigma$ and $r_c$. For charmonium these parameters are tuned to fit the masses of $\eta_c(1S)$, $\eta_c(2S)$, $J/\psi(1S)$ and $\chi_{c0}(1P)$, for bottomonium
these parameters are fixed by the masses of $\eta_b(1S)$, $\varUpsilon(1S)$, $\varUpsilon(2S)$ and $\chi_{b0}(1P)$. The parameters are listed in Table \ref{tab:parameters}.

\begin{table}[!ht]
\caption{\label{tab:parameters} Parameters used in our calculation.}
\begin{tabular}{c|c|c|c|c|c}
\hline
\makecell[c]{\vspace{1.0em}}\text{flavor}& $m/\text{GeV}$ & $\alpha_0$ & $b/\text{GeV}$ & $\sigma/\text{GeV}^2$ & $R_c/\text{fm}$ \\
\hline
c & 1.591 & 1.023& 0.144 & 1.30 & 0.375 \\
b & 4.997 & 0.875& 0.112 & 2.13 & 0.156 \\
\hline
\end{tabular}
\end{table}

The two-photon transition form factor of pseudoscalar meson is defined as \cite{Hoferichter2020}
\begin{eqnarray}\nonumber
&& \int d^4 x e^{iq_1\cdot x}\langle 0|T\{ j^\mu(x) j^\nu(0)\}|M\rangle \\\label{eq:TFF1S0}
&& = \epsilon^{\mu\nu\alpha\beta} q_{1\alpha}q_{2\beta}F_{P\gamma\gamma}(q_1^2,q_2^2).
\end{eqnarray}
The left hand side is the transition amplitude of the time-ordered currents between the meson state $|M\rangle$ and the vacuum $|0\rangle$. $j^\mu(x) =e_q \bar{\psi}(x)\gamma^\mu \psi(x)$, $e_q$ is the electric charge of the quark, $\psi(x)$ is the quark field operator. The right hand side expresses the Lorentz structure of the transition amplitude, $\epsilon^{\mu\nu\alpha\beta}$ is the Levi-Civita symbol, $q_{1\alpha}$ and $q_{2\beta}$ are the 4-momentum of the photons and $F_{P\gamma\gamma}(q_1^2,q_2^2)$ is the form factor.

Traditionally the meson is expressed by the mock meson state in quark model \cite{Godfrey1985,Lakhina2006,Sun2023}
\begin{eqnarray}\nonumber
 |M(p) \rangle &=& \sqrt{\frac{2E_p}{N_c}}\chi^{\bm{SM_S}}_{\bm{s\bar{s}}} \int\frac{d^3\bm{k} d^3\bm{\bar{k}}}{(2\pi)^3}\varphi_M\left(\bm{k}_r\right) \\\label{eq:oldMockMeson}
 && \cdot\delta^{(3)}(\bm{k}+\bm{\bar{k}}-\bm{p}) b^\dag_{\bm{ks}} d^\dag_{\bm{\bar{k}\bar{s}}}| 0 \rangle,
 \end{eqnarray}
where $\bm{p}$ and $E_p =\sqrt{M^2 + \bm{p}^2}$ are the momentum and energy of the meson, $M$ is the meson mass, $N_c$ is the color number. $b^\dag_{\bm{ks}}$ and $d^\dag_{\bm{\bar{k}\bar{s}}}$ are the creation operator of the quark and antiquark respectively. $\chi^{\bm{SM_S}}_{\bm{s\bar{s}}}$ is the spin wave function, with $\bm{S}$ being the total spin and $\bm{M_S}$ its z-projection. $\varphi_M\left(\bm{k}_r\right)$ is the wave function in momentum space, $\bm{k}_r=\frac{\bar{m}\bm{k} - m\bm{\bar{k}}}{m+\bar{m}}$ is the relative momentum between the quark and antiquark. Herein we use the same symbol $\varphi$ for the space wave function and the radial wave function, which could be distinguished by the arguments. In equation (\ref{eq:oldMockMeson}) a Clebsch-Gordan coefficient is indicated to combine the spin and orbital angular momentum into a total angular momentum of the meson.

The relativized mock meson state in analogy to Eq. (\ref{eq:oldMockMeson}) is
\begin{eqnarray}\nonumber
|M(p) \rangle &=& \sqrt{\frac{2E_p}{N_c}} \sum_{\bm{s,\bar{s}}} \int\frac{d^3\bm{k} d^3\bm{\bar{k}}}{(2\pi)^6}\frac{1}{\sqrt{2E_{\bm{k}}}} \frac{1}{\sqrt{2\bar{E}_{\bar{\bm{k}}}}}\\\label{eq:RelativisedMockMeson}
&&\bar{u}(\bm{k},\bm{s}) \Gamma_M v(\bar{\bm{k}},\bar{\bm{s}}) b^\dag_{\bm{ks}} d^\dag_{\bm{\bar{k}\bar{s}}} | 0 \rangle.
\end{eqnarray}
$E_{\bm{k}}=\sqrt{m^2+\bm{k}^2}$ and $\bar{E}_{\bar{\bm{k}}} = \sqrt{\bar{m}^2 + \bar{\bm{k}}^2}$ are the kinetic energy of the quark and antiquark. ${u}(\bm{k},\bm{s})$ and $v(\bar{\bm{k}},\bar{\bm{s}})$ are the Dirac spinors of the quark and antiquark respectively. $\Gamma_M$ is the meson Bethe-Salpeter amplitude. For pseudoscalar meson it's generally composed of 4 terms \cite{Chen2017}. Herein we only employ the main term for simplicity, i.e.
\begin{equation}\label{eq:BSA1S0}
 \Gamma_P=\Gamma_P(k_r,p) = \gamma_5 \phi_P(k_r,p),
\end{equation}
where $k_r$ and $p$ are the relative and total 4-momentum, $\gamma_5$ is Dirac matrix, $\phi_P(k_r,p)$ is the scalar wave function. Analogically, for scalar meson
\begin{equation}\label{eq:BSA3P0}
 \Gamma_S=\Gamma_S(k_r,p) = \bm{1}_{4\times 4} \phi_S(k_r,p),
\end{equation}
where $\bm{1}_{4\times 4}$ is the unit matrix in Dirac space, and $\phi_S(k_r,p)$ is the scalar wave function.

\begin{figure}[!t]
\centering
 \includegraphics[width=0.48\textwidth]{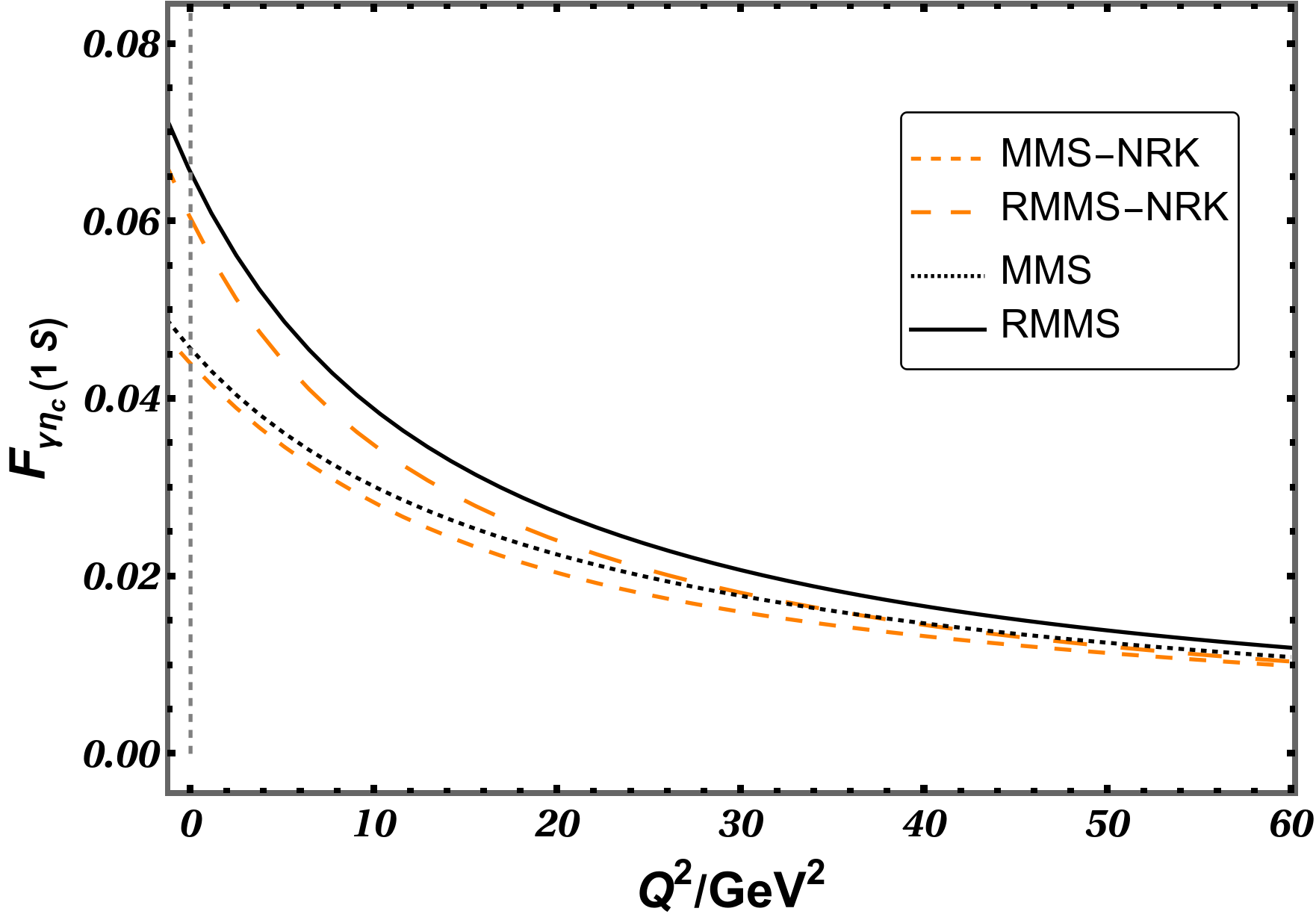}
 \caption{\label{fig:tffetacnrT} (colored online) The two-photon transition form factor of $\eta_c(1S)$. The black solid line (RMMS) is the result with relativized mock meson state, Eq. (\ref{eq:RelativisedMockMeson}) and the relativistic kinetic energy, Eq. (\ref{eq:RelKEnergy}). The orange long-dashed line (RMMS-NRK) is the result with relativized mock meson state, Eq. (\ref{eq:RelativisedMockMeson}) and the non-relativistic kinetic energy, Eq. (\ref{eq:nRelKEnergy}). The black dotted line (MMS) is the result with the old mock meson state, Eq. (\ref{eq:oldMockMeson}) and the relativistic kinetic energy, Eq. (\ref{eq:RelKEnergy}). The orange short-dashed line (MMS-NRK) is the result with old mock meson state, Eq. (\ref{eq:oldMockMeson}) and the non-relativistic kinetic energy, Eq. (\ref{eq:nRelKEnergy}).}
\end{figure}

There are several reasons to go from Eq. (\ref{eq:oldMockMeson}) to Eqs. (\ref{eq:RelativisedMockMeson})-(\ref{eq:BSA3P0}):
\begin{enumerate}[label=\roman*)]
 \item Eq. (\ref{eq:oldMockMeson}) is not Lorentz covariant, so it leads to systematic errors in describing decay constants and form factors \cite{Lakhina2006,Sun2023}. Conversely, the meson state described by Eqs. (\ref{eq:RelativisedMockMeson})-(\ref{eq:BSA3P0}) is formally Lorentz covariant.
 \item Compared with the non-relativistic form, the relativistic current $j^\mu(x) = e_q \bar{\psi}(x)\gamma^\mu \psi(x)$ leads to substantial corrections to the decay constant and sigle quark form factor \cite{Lakhina2006}. We expect that the relativistic mock meson state also have a significant impact on the form factors.
 \item \label{item:relativistic} Eqs. (\ref{eq:RelativisedMockMeson})-(\ref{eq:BSA3P0}) introduce relativistic corrections from two aspects, one is from the Dirac spinor, the other is from the wave function. To assess the relativistic correction from the wave function, $\phi_P(k_r,p)$ and $\phi_S(k_r,p)$ should be solved from a covariant dynamical equation, e.g., the Bethe-Salpeter equation, which is technically challenging. However, signs indicate that the relativistic correction from the Dirac spinor is major and that from the wavefunction is minor. We calculated the effect of the kinetic energy on the wave function, i.e., we compared results from Eq. (\ref{eq:RelKEnergy}) and Eq. (\ref{eq:nRelKEnergy}). From Fig. \ref{fig:tffetacnrT} we see that the relativistic correction from the Dirac spinor is dominant compared to that from the wavefunction. We expect other corrections to the wave fuction to be small, though the accurate wavefunction is still an open question.
\end{enumerate}

 Considering the ``\ref{item:relativistic}" point, we use the relativistic kinetic energy, Eq. (\ref{eq:RelKEnergy}), in the following calculation, and we won't solve $\phi_P(k_r,p)$ and $\phi_S(k_r,p)$ rigorously. We relate $\phi_P(k_r,p)$ and $\phi_S(k_r,p)$ to the non-relativistic wave functions by requiring that Eq. (\ref{eq:RelativisedMockMeson}) approaches Eq. (\ref{eq:oldMockMeson}) in the non-relativistic limit. The derivations are put in \ref{sec:appendixA}, and the results are \begin{eqnarray}\label{eq:Relate1S0}
\phi_P(k_r,p)\!\! &=&\!\! \frac{\varphi_P\left(|\bm{k}_r|\right)}{\sqrt{8\pi}} (2\pi)^3\delta^{(3)}(\bm{k}+\bm{\bar{k}}-\bm{p}),\\\label{eq:Relate3P0}
 \phi_S(k_r,p)\!\! &=&\!\!\! \frac{2E_{\bm{k}} \bar{E}_{\bar{\bm{k}}}}{E_{\bm{k}} + \bar{E}_{\bar{\bm{k}}}}\frac{\varphi_S\left(|\bm{k}_r|\right)}{\sqrt{8\pi}|\bm{k}_r|} (2\pi)^3\delta^{(3)}(\bm{k}\!+\!\bm{\bar{k}}\!-\!\bm{p}).                                                                                                                                                                                                                                                                                                                                                                                                                                                                                                                                                                                                                                                        \end{eqnarray}
 Eqs. (\ref{eq:RelativisedMockMeson})-(\ref{eq:Relate3P0}) represent relativized mock meson states, where the Dirac spinors import kinetic relativistic corrections and the dynamic wave functions are the same as Eq. (\ref{eq:oldMockMeson}). Note that Eqs. (\ref{eq:Relate1S0}) and (\ref{eq:Relate3P0}) are not Lorentz invariant, so these relativized mock meson states still show a little frame dependence, but this dosen't affect our arguments in this paper.

 We work in the meson static frame to calculate the two-photon transition form factor, where the 4-momentum of the meson and the two photons are
\begin{eqnarray}
 p &=& (M, \vec{0}),\\
 q_1 &=& (\frac{M^2+Q_1^2 - Q_2^2}{2M},0,0,\lambda),\\
 q_2 &=& (\frac{M^2-Q_1^2 + Q_2^2}{2M},0,0,-\lambda).\end{eqnarray}
$\lambda = \frac{1}{2M}\{ M^4+Q_1^4+Q_2^4 -2(M^2Q_1^2 + M^2Q_2^2 + Q_1^2Q_2^2) \}^{1/2}$, and $Q_1^2 = q_1^2$, $Q_2^2 = q_2^2$.

Put Eq. (\ref{eq:oldMockMeson}) into Eq. (\ref{eq:TFF1S0}), we get the pseudoscalar meson two-photon transition form factor corresponding to the old mock meson state. Put Eq. (\ref{eq:RelativisedMockMeson}) into Eq. (\ref{eq:TFF1S0}), we get the expression corresponding to the relativized mock meson state. In this paper, we focus on the single-tag form factor in the space-like region, $F_{\gamma P}(Q^2) = F_{P\gamma\gamma}(-Q^2,0)$. The derivation and the expressions of the form factors, and those analogs of the scalar meson are all put in \ref{sec:appendixB}.

\section{The Charmonium}\label{sec:resultsCC}

\subsection{Charmonium spectrum and wave function}\label{subsec:spectrumCC}

\begin{table}[th!]
 \caption{\label{tab:mcc} Mass spectrum of charmonium (in GeV). $M^{\textmd{QM}}_{c\bar{c}}$ is our result via quark model, with the parameters listed in Table \ref{tab:parameters}. Note that the parameters are tuned to fit the observed masses of $\eta_c(1S)$, $\eta_c(2S)$, $J/\psi(1S)$ and $\chi_{c0}(1P)$. $M^{\textmd{expt.}}_{c\bar{c}}$ is the experimental value \cite{Navas2024}. }
\begin{tabular}{c|c|c|l|l}
\hline
\makecell[c]{\vspace{1.0em}}$n^{2S+1}L_J$&state &$J^{\textmd{PC}}$&  $M^{\textmd{QM}}_{c\bar{c}}$	& $M^{\textmd{expt.}}_{c\bar{c}}$\\
\hline
$1^1S_0$& $\eta_c(1S)$& $0^{-+}$  &  2.984 (input) &  2.984(0.4) \\
$2^1S_0$& $\eta_c(2S)$& $0^{-+}$  &  3.638 (input) &  3.638(1) \\
$3^1S_0$& $\eta_c(3S)$& $0^{-+}$  &   4.048 &\phantom{0.}--\\
$1^3S_1$& $J/\psi(1S)$& $1^{--}$  &   3.097 (input) &  3.097(0) \\
$2^3S_1$& $\psi(2S)$& $1^{--}$ &     3.684 & 3.686(0.1) \\
$3^3S_1$& $\psi(4040)$& $1^{--}$ &   4.079 &  4.039(1) \\
$1^3P_0$& $\chi_{c0}(1P)$& $0^{++}$  & 3.415 (input) &  3.415(0.3) \\
$2^3P_0$& $\chi_{c0}(2P)$& $0^{++}$ & 3.894 &  \phantom{0.}-- \\
$3^3P_0$& $\chi_{c0}(3P)$& $0^{++}$ & 4.251 &  \phantom{0.}-- \\
$1^1P_1$& $h_{c}(1P)$& $1^{+-}$  & 3.510 &  3.525(0.1) \\
$2^1P_1$& $h_{c}(2P)$& $1^{+-}$  & 3.933 &  \phantom{0.}-- \\
$3^1P_1$& $h_{c}(3P)$& $1^{+-}$  &  4.271 &  \phantom{0.}-- \\
$1^3P_1$& $\chi_{c1}(1P)$& $1^{++}$ & 3.510 &  3.511(0.1) \\
$2^3P_1$& $\chi_{c1}(2P)$& $1^{++}$ & 3.943 &  \phantom{0.}-- \\
$3^3P_1$& $\chi_{c1}(3P)$& $1^{++}$ & 4.284 &  \phantom{0.}-- \\
$1^3P_2$& $\chi_{c2}(1P)$& $2^{++}$ & 3.561 &  3.556(0.1) \\
$2^3P_2$& $\chi_{c2}(3930)$& $2^{++}$ & 3.971&  3.923(1) \\
$3^3P_2$& $\chi_{c2}(3P)$& $2^{++}$  & 4.302 &  -- \\
\hline
\end{tabular}
\end{table}

\begin{figure}[!th]
\centering
 \includegraphics[width=0.48\textwidth]{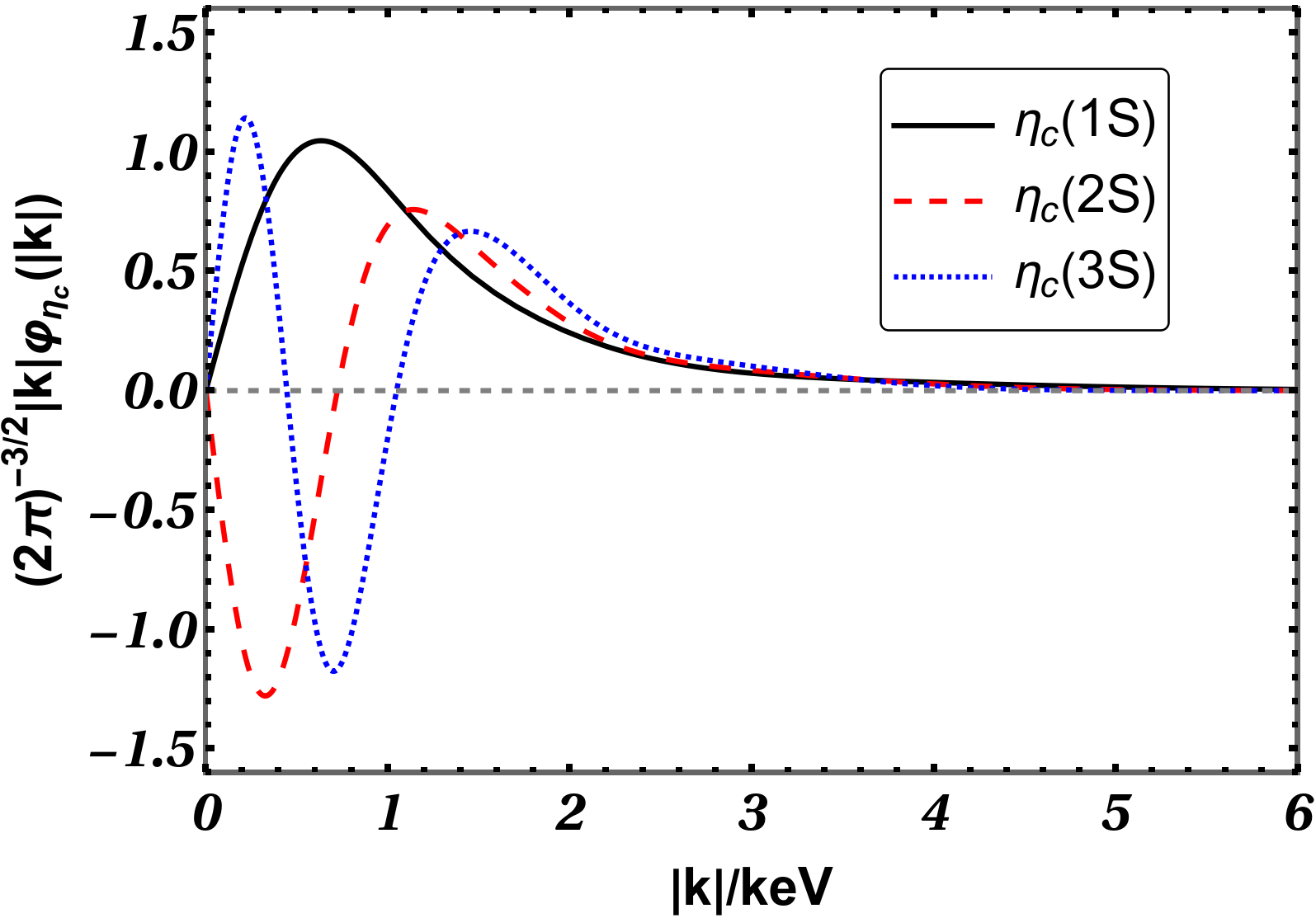}
 \includegraphics[width=0.48\textwidth]{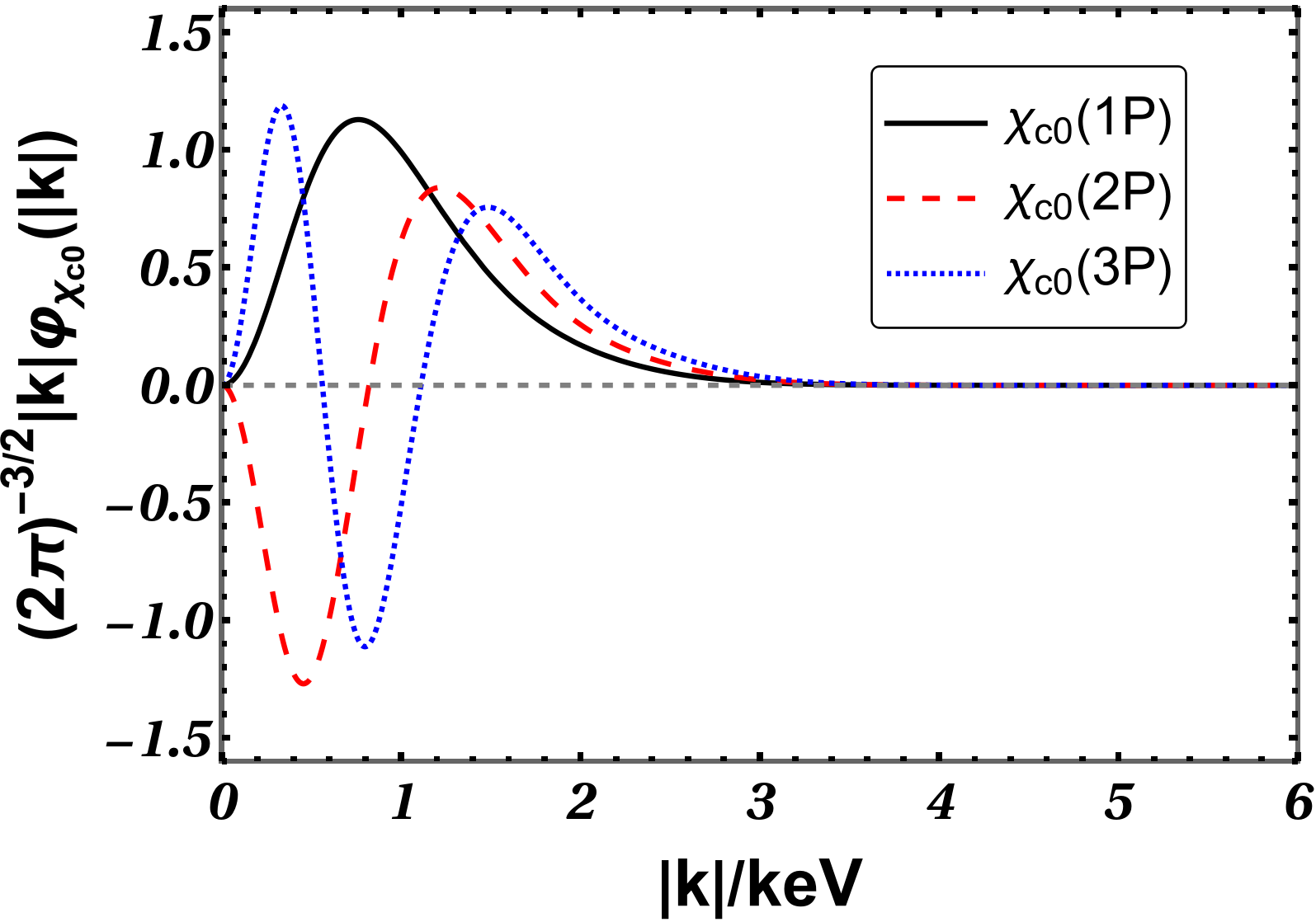}
 \caption{\label{fig:wfcc} (colored online) The radial wave function of $\eta_c(nS)$ (upper) and $\chi_{c0}(nP)$ (lower).}
\end{figure}

The charmonium spectrum is listed in Table \ref{tab:mcc}. Note that the interaction parameters are fitted by the observed masses of $\eta_c(1S)$, $\eta_c(2S)$, $J/\psi(1S)$ and $\chi_{c0}(1P)$, i.e., these four masses are inputs of our model, and all the other masses and the two-photon transition form factors are predictions. For the mesons under the $D\bar{D}$ threshold, the predicted masses agree with the observed values quite well. For those above the $D\bar{D}$ threshold, e.g., $\psi(4040)$ and $\chi_{c2}(3930)$, the deviation is about $40\sim50 \text{ MeV}$. Coupled-channel effect is needed for a more precise prediction \cite{Deng2024}. In this paper, we focus on the $\eta_c(nS)$ and $\chi_{c0}(nP)$ for $n =1,2,3$, among which $\eta_c(3S)$, $\chi_{c0}(2P)$ and $\chi_{c0}(3P)$ are above the $D\bar{D}$ threshold. For a first step study of the effect of the relativized mock meson state, we ignore coupled-channel effect herein.

The radial wavefunctions of $\eta_c(nS)$ and $\chi_{c0}(nP)$ are in Fig. \ref{fig:wfcc}. The radial wave function is normalized as
\begin{equation}
 \int_0^\infty \frac{d|\bm{k}|}{(2\pi)^3}\bm{k}^2\left|\varphi_M(|\bm{k}|)\right|^2 =1.
\end{equation}
The same wave functions are used both in the old mock meson state, Eq.(\ref{eq:oldMockMeson}), and the relativized mock meson state, Eqs. (\ref{eq:RelativisedMockMeson})-(\ref{eq:Relate3P0}). So the relativized mock meson state only takes into account the kinetic relativistic correction imported by the Dirac spinors and the dynamic wave function is not modified. Our results in the following show that this kinetic relativistic correction is crucial and dominant in studying the two-photon transition form factors.

\subsection{The two-photon transition form factor and decay width}\label{subsec:formfactorCC}

\begin{figure}[!t]
\centering
 \includegraphics[width=0.48\textwidth]{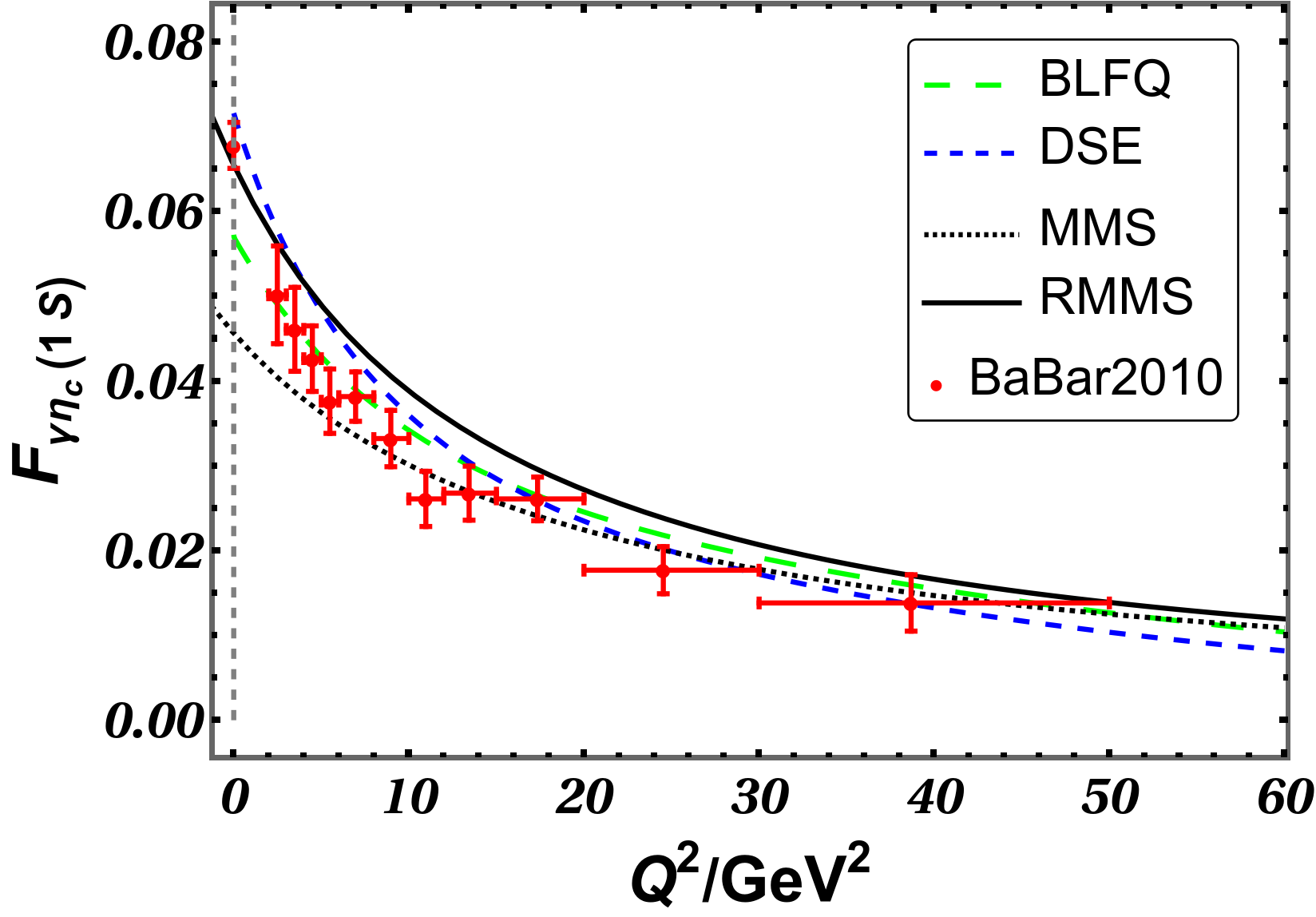}
 \includegraphics[width=0.48\textwidth]{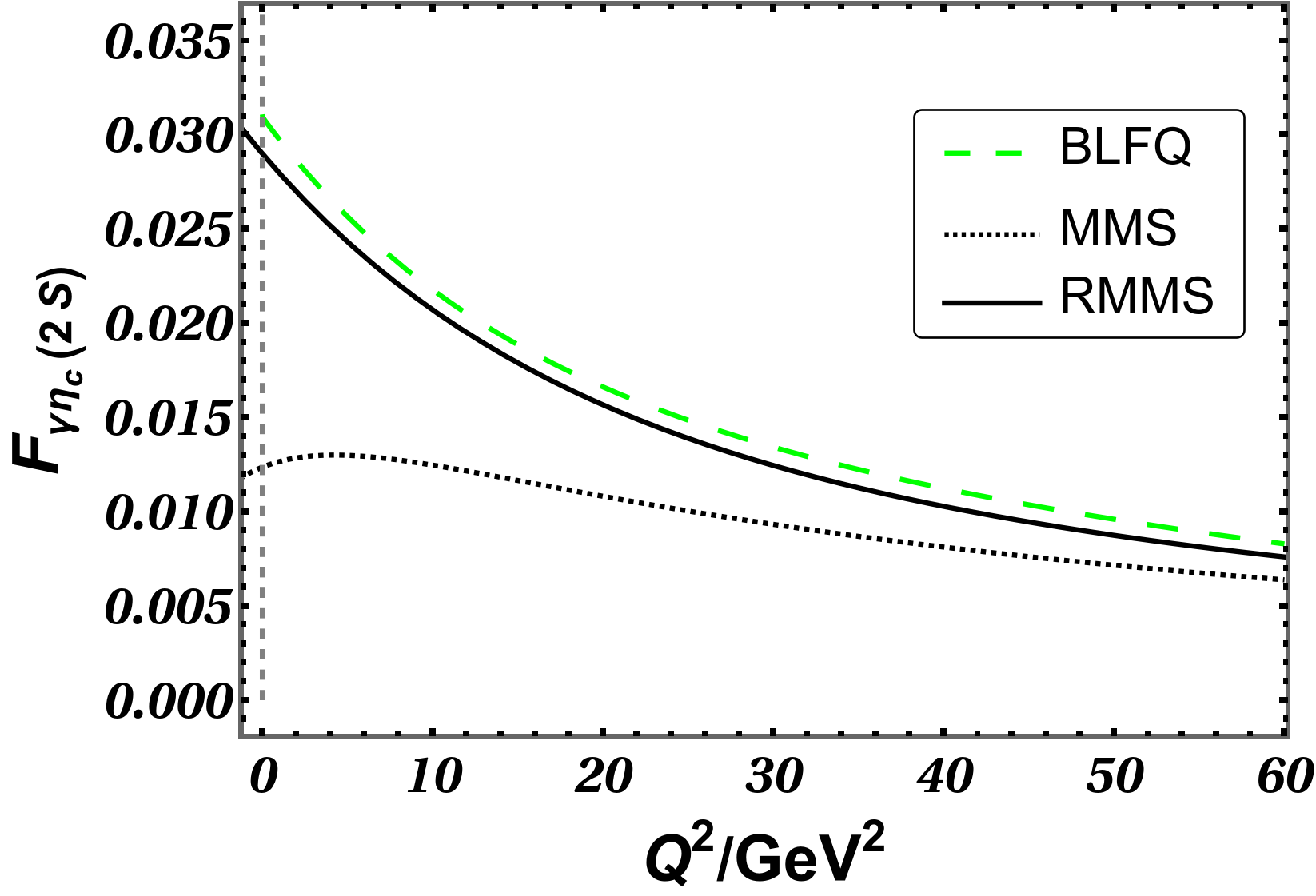}
 \includegraphics[width=0.48\textwidth]{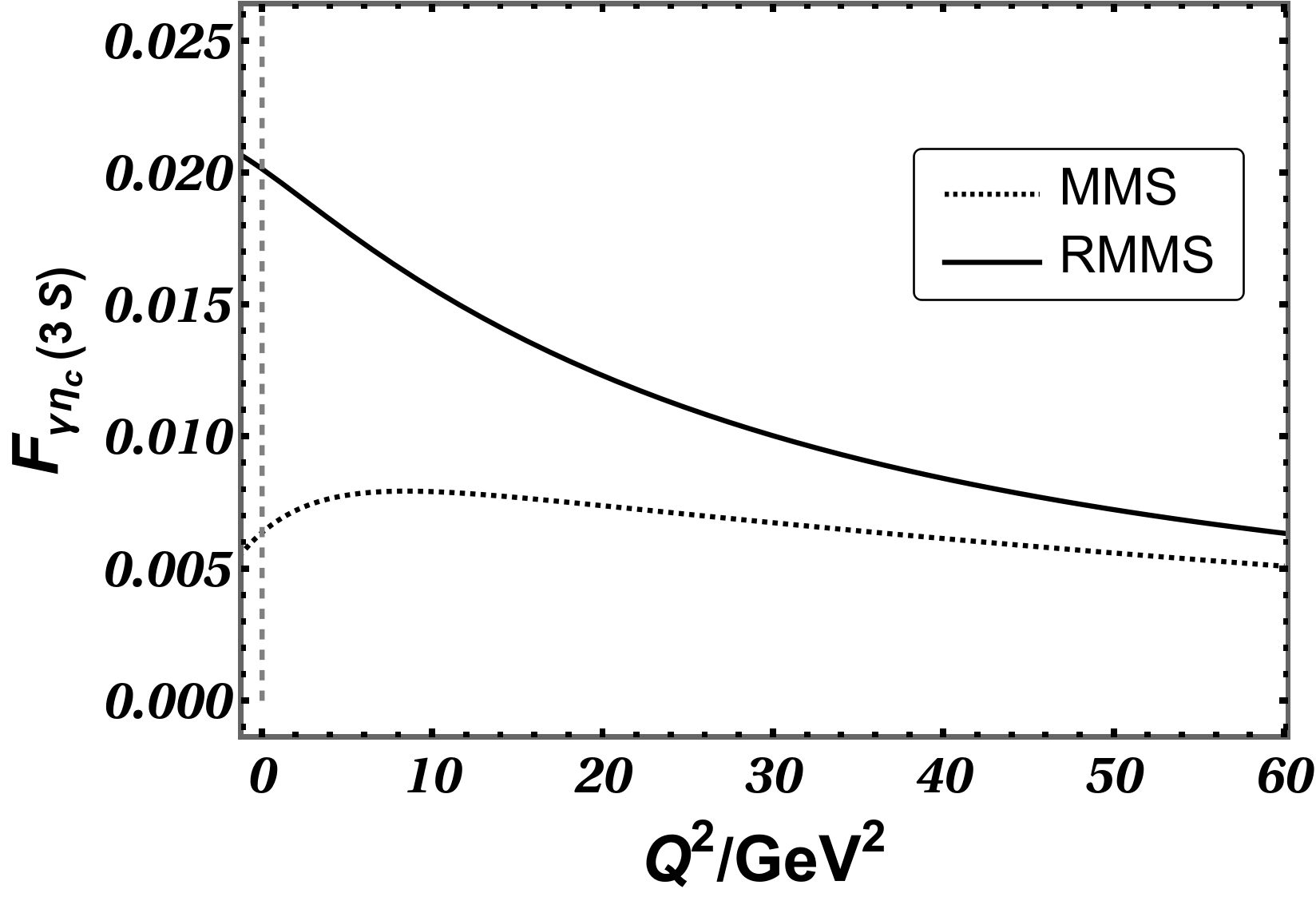}
 \caption{\label{fig:tffetac} (colored online) The two-photon transition form factor of $\eta_c(1S)$ (upper), $\eta_c(2S)$ (middle) and $\eta_c(3S)$ (lower). The black solid lines are our results with relativized mock meson state, Eq. (\ref{eq:RelativisedMockMeson}). The black dotted lines are our results with the old mock meson state, Eq. (\ref{eq:oldMockMeson}). The blue short-dashed line is the Dyson-Schwinger equation result \cite{Chen2017}. The green long-dashed line is the result of the basis light front quantization approach \cite{Li2022}. The red dots are the BaBar results \cite{Lees2010}.}
\end{figure}

\begin{figure}[!t]
\centering
 \includegraphics[width=0.48\textwidth]{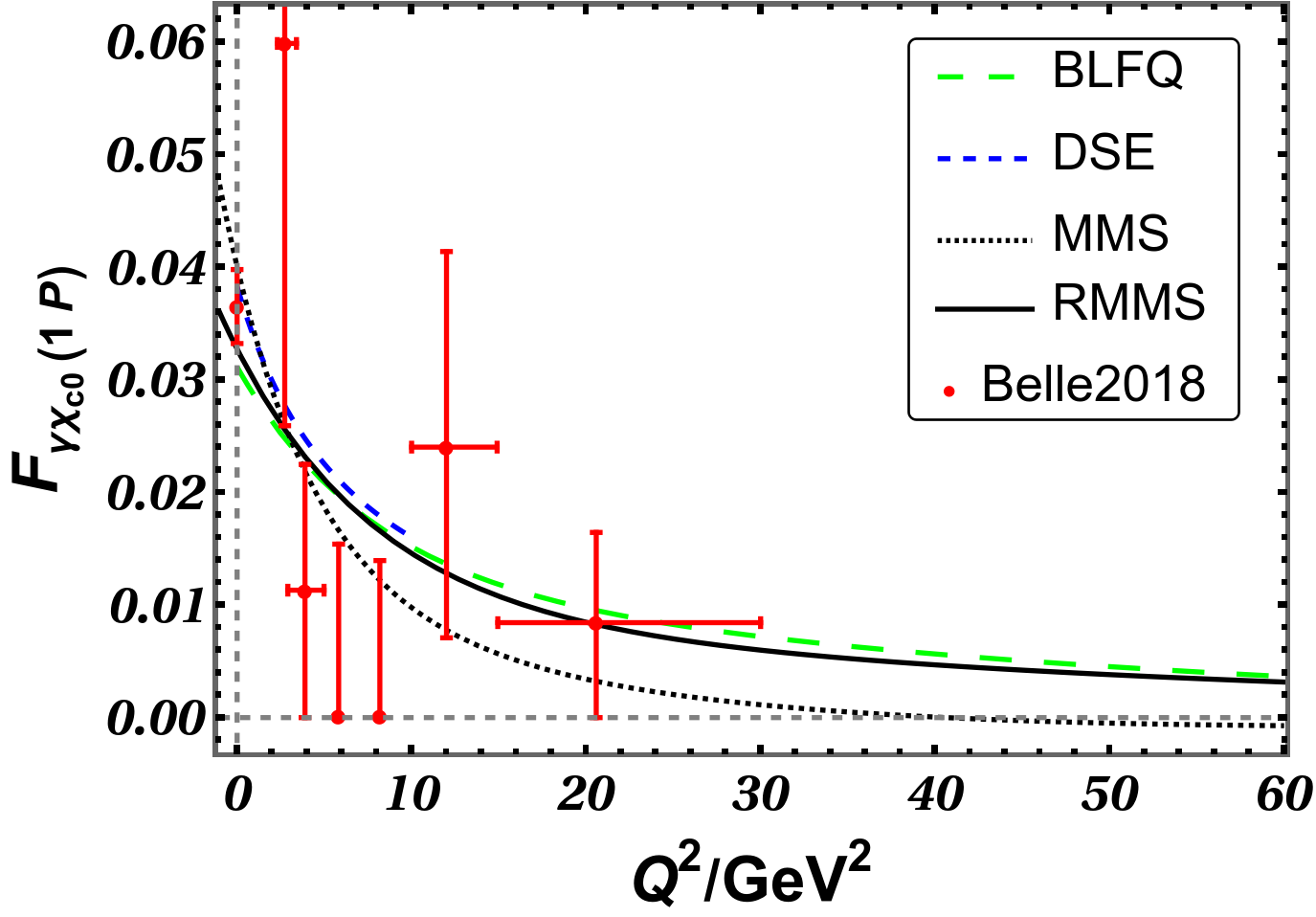}
 \includegraphics[width=0.48\textwidth]{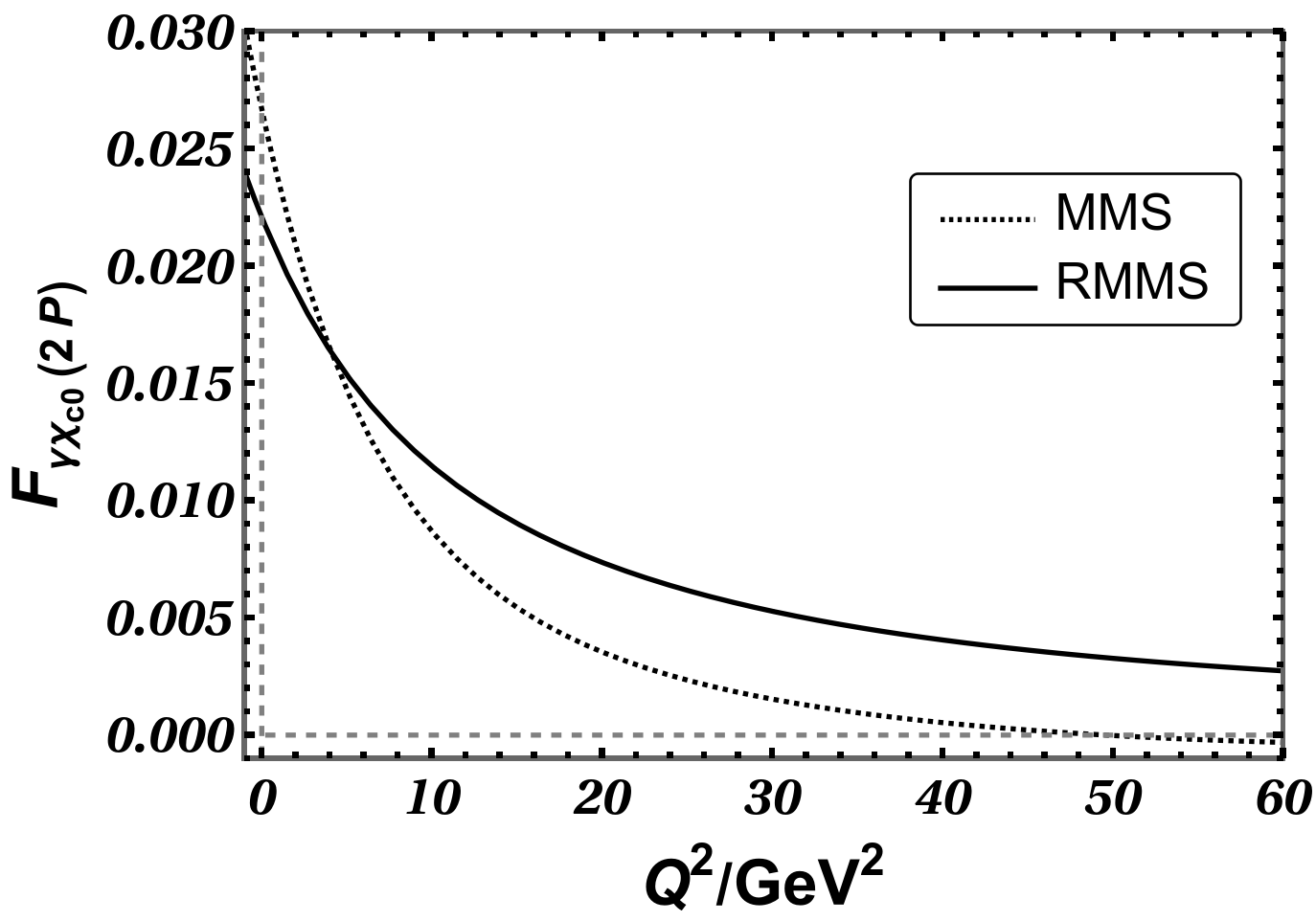}
 \includegraphics[width=0.48\textwidth]{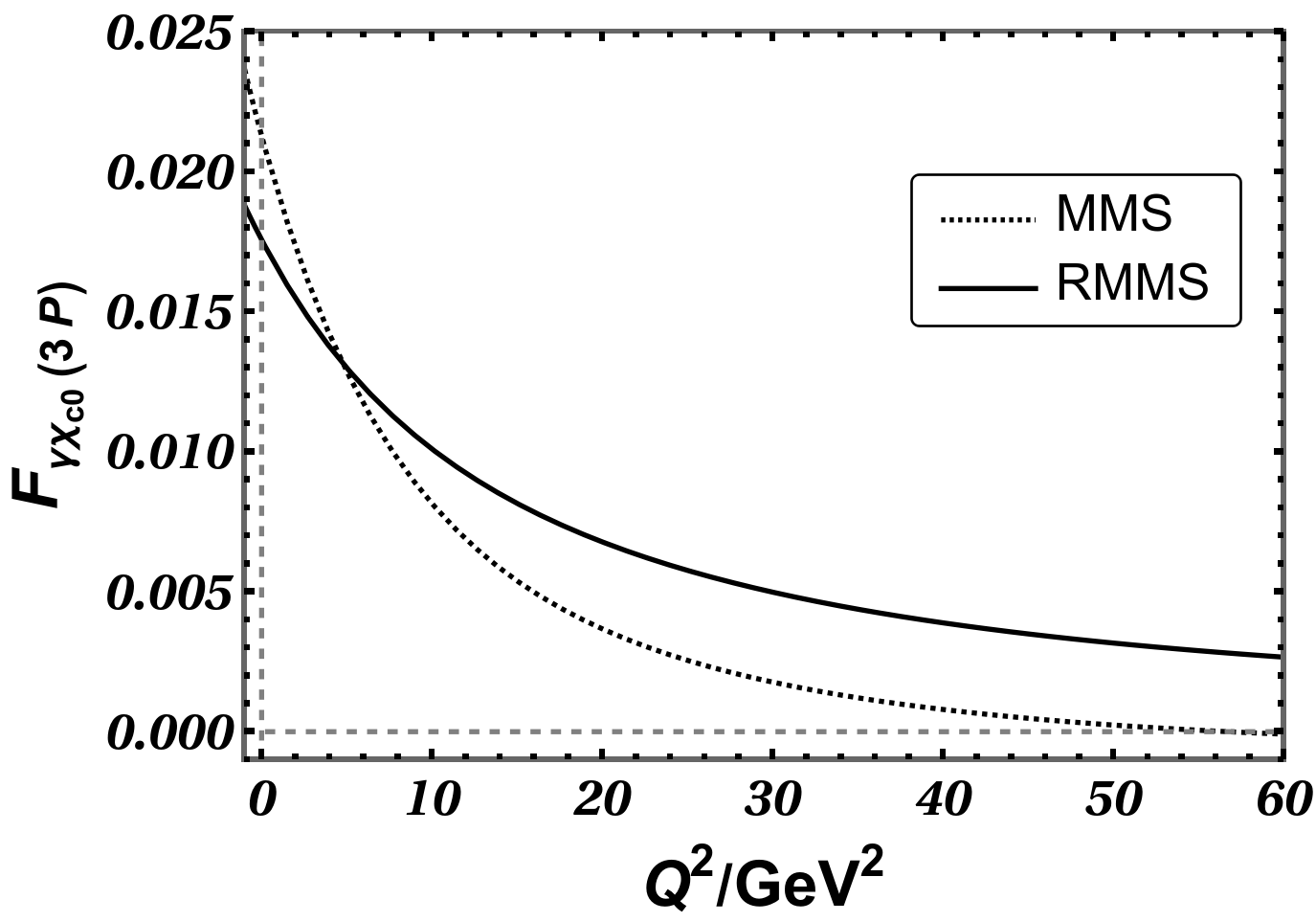}
 \caption{\label{fig:tffChic0} (colored online) The two-photon transition form factor of $\chi_{c0}(1P)$ (upper), $\chi_{c0}(2P)$ (middle) and $\chi_{c0}(3P)$ (lower). The black solid lines are our results with relativized mock meson state, Eq. (\ref{eq:RelativisedMockMeson}). The black dotted lines are our results with the old mock meson state, Eq. (\ref{eq:oldMockMeson}). The blue short-dashed line is the Dyson-Schwinger equation result \cite{Chen2017}. The green long-dashed line is the result of the basis light front quantization approach \cite{Li2022}. The red dots are the preliminary Belle results \cite{Masuda2018}.}
\end{figure}

\begin{table*}[t!]
 \caption{\label{tab:DecayWidthcc} Two photon decay width of charmonium (in keV). $\Gamma^{\textmd{MMS}}_{\gamma\gamma}$ is our result with the old mock state, $\Gamma^{\textmd{RMMS}}_{\gamma\gamma}$ is our result with relativized mock state. $\Gamma^{\textmd{expt.}}_{\gamma\gamma}$ is the experimental value \cite{Navas2024}. $\Gamma^{\textmd{BLFQ}}_{\gamma\gamma}$ is the basis light-front quantization result \cite{Li2022}. $\Gamma^{\textmd{DSE}}_{\gamma\gamma}$ is the Dyson-Schwinger equation result \cite{Chen2017}. $\Gamma^{\textmd{LQCD}}_{\gamma\gamma}$ are the lattice QCD results \cite{Dudek2006,Meng2023}. $\Gamma^{\textmd{NRQM}}_{\gamma\gamma}$ is a non-relativistic quark model result \cite{Lakhina2006}. $\Gamma^{\textmd{RQM}}_{\gamma\gamma}$ is a relativistic quark model result \cite{Ebert2003}. }
\begin{tabular}{c|c|c|c|c|c|c|c|c|c}
\hline
\makecell[c]{\vspace{1.0em}}&  $\Gamma^{\textmd{MMS}}_{\gamma\gamma}$&  $\Gamma^{\textmd{RMMS}}_{\gamma\gamma}$& $\Gamma^{\textmd{expt.}}_{\gamma\gamma}$\cite{Navas2024}	& $\Gamma^{\textmd{BLFQ}}_{\gamma\gamma}$	\cite{Li2022}& $\Gamma^{\textmd{DSE}}_{\gamma\gamma}$ \cite{Chen2017}& $\Gamma^{\textmd{LQCD}}_{\gamma\gamma}$ \cite{Dudek2006} & $\Gamma^{\textmd{LQCD}}_{\gamma\gamma}$ \cite{Meng2023}  & $\Gamma^{\textmd{NRQM}}_{\gamma\gamma}$ \cite{Lakhina2006} & $\Gamma^{\textmd{RQM}}_{\gamma\gamma}$ \cite{Ebert2003}\\
\hline
 $\eta_c(1S)$&  2.35 & 4.84 &$5.1\pm 0.4$& 3.7(6) &6.39  & 2.65(99) & 6.67(17) & 7.18 & 5.5  \\
 $\eta_c(2S)$&  0.32 & 1.68  &$2.1\pm 1.4$& 1.9(4) & & & & 1.71& 1.8  \\
 $\eta_c(3S)$&  0.11 & 1.13 &--&  & & & & 1.21 &  \\
 $\chi_{c0}(1P)$& 2.84 & 1.92 & $2.2\pm 0.2$& 1.7(4) & 2.39 & 2.41(1.04) & & 3.28 & 2.9  \\
 $\chi_{c0}(2P)$& 1.80 & 1.25 &-- & 0.68(22) & & & & &  1.9  \\
 $\chi_{c0}(3P)$& 1.42 & 1.00&--  &  & & & & & \\
\hline
\end{tabular}
\end{table*}

The prediction of two-photon transition form factors of $\eta_c(nS)$ are in Fig. \ref{fig:tffetac} and those of $\chi_{c0}(nP)$ are in Fig. \ref{fig:tffChic0}. Results are easily readable from the figures. In the following we will comment on them systematically.

First, let's focus on $F_{\gamma\eta_c}(1S)$, the upper panel of Fig. \ref{fig:tffetac}. The black dotted line is our result with the old mock meson state (MMS), and the black solid line is our result with relativized mock meson state (RMMS). They approach each other in the ultraviolet region, and deviate remarkably in the infrared region.  In the lower $Q^2$ region ($Q^2 \lesssim 10 \text{ GeV}$), the RMMS result is more consistent with the Dyson-Schwinger equation (DSE) result \cite{Chen2017}, the blue short-dashed line, and the BaBar experiment results, the red dots with error bars \cite{Masuda2018}. Although the result of the basis light front quantization (BLFQ) approach \cite{Li2022}, the green long-dashed line is also consistent with the Babar result, the error at $Q^2=0$ is considerable. The RMMS and DSE results matches the experimental value at $Q^2=0$ much better.

 In the higher $Q^2$ region ($ 40 \text{ GeV} \lesssim Q^2 \lesssim 60 \text{ GeV}$), the MMS, RMMS and BLFQ results approach each other, and they are higher than the DSE result. This indicates that the quark model results approach a different $Q^2 \to \infty$ asymptotic value compared with the DSE result, which is manifestly visible from Fig. 7 of Ref. \cite{Li2022}. As a low energy effective theory, the  asymptotic behavior of quark model is not strictly constrained, so the deviation in the higher $Q^2$ region should not be taken too seriously. In the following, we will focus on the lower $Q^2$ region.

The figure of $F_{\gamma\eta_c}(1S)$ indicates that the kinetic relativistic correction imported by the Dirac spinors, in Eq.(\ref{eq:RelativisedMockMeson}), is crucial and dominant in studying the two-photon transition form factors, especially in the infrared region. This point is further supported by $F_{\gamma\eta_c}(2S)$, the middle panel of Fig. \ref{fig:tffetac}, and $F_{\gamma\chi_{c0}}(1P)$, the upper panel of Fig. \ref{fig:tffChic0}. For $F_{\gamma\eta_c}(2S)$, RMMS result is consistent with the BLFQ result, while the MMS result declines significantly in the infrared region. For $F_{\gamma\chi_{c0}}(1P)$, RMMS result is consistent with the DSE and BLFQ result, while the MMS result decreases faster and declines to negative value in the ultraviolet region. RMMS result also matches the experimental value $F_{\gamma\chi_{c0}}(1P)(0)$ better than MMS result. A preliminary experiment result of $F_{\gamma\chi_{c0}}(1P)(Q^2)$ \cite{Masuda2018} is also plotted, but it's senseless to be compared with due to the large errors.

Overall, the RMMS results are consistent with other theoretical predictions and the available experimental values. The form factor $F_{\gamma\eta_c}(3S)$ is predicted in Fig. \ref{fig:tffetac}, $F_{\gamma\chi_{c0}}(2P)$ and $F_{\gamma\chi_{c0}}(3P)$ are predicted in Fig. \ref{fig:tffChic0}.

Our predictions of two photon decay width of $F_{\gamma\eta_c}(nS)$ and $F_{\gamma\chi_{c0}}(nP)$
are listed in Table \ref{tab:DecayWidthcc}. In general, RMMS result matches the experimental values, BLFQ result and DSE result much better than the MMS result. The lattice QCD result varies from one another \cite{Dudek2006,Meng2023}. $\Gamma^{\textmd{NRQM}}_{\gamma\gamma}$ \cite{Lakhina2006} is the result from combination of non-relativistic quark model and an effective Lagrange. $\Gamma^{\textmd{RQM}}_{\gamma\gamma}$ \cite{Ebert2003}, the result from a relativistic quark model is also consistent with our RMMS result.

\section{The Bottomonium}\label{sec:resultsBB}

\subsection{Bottomonium spectrum and wave function}\label{subsec:spectrumBB}

\begin{table}[!t]
 \caption{\label{tab:mbb} Mass spectrum of bottomonium (in GeV). $M^{\textmd{QM}}_{b\bar{b}}$ is our result via quark model, with the parameters listed in Table \ref{tab:parameters}. Note that the parameters are tuned to fit the masses of $\eta_b(1S)$, $\varUpsilon(1S)$, $\varUpsilon(2S)$ and $\chi_{b0}(1P)$. $M^{\textmd{expt.}}_{b\bar{b}}$ is the experimental value \cite{Navas2024}. }
\begin{tabular}{c|c|c|l|l}
\hline
\makecell[c]{\vspace{1.0em}}$n^{2S+1}L_J$&state &$J^{\textmd{PC}}$&  $M^{\textmd{QM}}_{b\bar{b}}$	& $M^{\textmd{expt.}}_{b\bar{b}}$\\
\hline
$1^1S_0$ & $\eta_b(1S)$& $0^{-+}$  & \phantom{0}9.399 (input) &  \phantom{0}9.399(2) \\
$2^1S_0$ & $\eta_b(2S)$& $0^{-+}$  & 10.005 &  \phantom{0}9.999(4) \\
$3^1S_0$ & $\eta_b(3S)$& $0^{-+}$  & 10.321 &  \phantom{00.}-- \\
$1^3S_1$ & $\varUpsilon(1S)$& $1^{--}$  &  \phantom{0}9.460 (input) & \phantom{0}9.460(0.3) \\
$2^3S_1$ & $\varUpsilon(2S)$& $1^{--}$  & 10.023 (input) & 10.023(0.3) \\
$3^3S_1$ & $\varUpsilon(3S)$& $1^{--}$  & 10.333 & 10.355(1) \\
$4^3S_1$ & $\varUpsilon(4S)$& $1^{--}$  & 10.566 & 10.579(1) \\
$1^3P_0$ & $\chi_{b0}(1P)$& $0^{++}$  & \phantom{0}9.859 (input) & \phantom{0}9.859(1) \\
$2^3P_0$ & $\chi_{b0}(2P)$& $0^{++}$  & 10.216 & 10.233(1) \\
$3^3P_0$ & $\chi_{b0}(3P)$& $0^{++}$  & 10.472 & \phantom{00.}-- \\
$1^1P_1$ & $h_{b}(1P)$ & $1^{+-}$  &  \phantom{0}9.915 &  \phantom{0}9.899(1) \\
$2^1P_1$ & $h_{b}(2P)$ & $1^{+-}$  & 10.245 &  10.260(1) \\
$3^1P_1$ & $h_{b}(3P)$ & $1^{+-}$  & 10.491 &  \phantom{00.}-- \\
$1^3P_1$ & $\chi_{b1}(1P)$& $1^{++}$  & \phantom{0}9.907 & \phantom{0}9.893(1) \\
$2^3P_1$ & $\chi_{b1}(2P)$& $1^{++}$  & 10.242 & 10.255(1) \\
$3^3P_1$ & $\chi_{b1}(3P)$& $1^{++}$  & 10.490 & 10.513(1) \\
$1^3P_2$ & $\chi_{b2}(1P)$& $2^{++}$  & \phantom{0}9.938 & \phantom{0}9.912(1) \\
$2^3P_2$ & $\chi_{b2}(2P)$& $2^{++}$  & 10.261 & 10.269(1) \\
$3^3P_2$ & $\chi_{b2}(3P)$& $2^{++}$  & 10.503 & 10.524(1) \\
$1^3D_2$ & $\varUpsilon_2(1D)$& $2^{--}$ &  10.144 & 10.164(1) \\
\hline
\end{tabular}
\end{table}

\begin{table}[t!]
 \caption{\label{tab:DecayWidthbb} Two photon decay width of bottomonium (in keV). $\Gamma^{\textmd{MMS}}_{\gamma\gamma}$ is our result with the old mock state, $\Gamma^{\textmd{RMMS}}_{\gamma\gamma}$ is our result with relativized mock state. $\Gamma^{\textmd{DSE}}_{\gamma\gamma}$ is the Dyson-Schwinger equation result \cite{Chen2017}. $\Gamma^{\textmd{NRQM}}_{\gamma\gamma}$ is a non-relativistic quark model result \cite{Lakhina2006}. $\Gamma^{\textmd{RQM}}_{\gamma\gamma}$ is a relativistic quark model result \cite{Ebert2003}. $\Gamma^{\textmd{HQSS}}_{\gamma\gamma}$ is an estimatation from heavy-quark spin-symmetry \cite{Lansberg2007}.}
\begin{tabular}{c|c|c|c|c|c|c}
\hline
\makecell[c]{\vspace{2em}}&  $\Gamma^{\textmd{MMS}}_{\gamma\gamma}$&  $\Gamma^{\textmd{RMMS}}_{\gamma\gamma}$	& \makecell[c]{$\Gamma^{\textmd{DSE}}_{\gamma\gamma}$\\ \cite{Chen2017}}& \makecell[c]{$\Gamma^{\textmd{NRQM}}_{\gamma\gamma}$\\ \cite{Lakhina2006}} & \makecell[c]{$\Gamma^{\textmd{RQM}}_{\gamma\gamma}$\\ \cite{Ebert2003}}& \makecell[c]{$\Gamma^{\textmd{HQSS}}_{\gamma\gamma}$\\ \cite{Lansberg2007}}\\
\hline
 $\eta_b(1S)$&  0.237 & 0.354 & 0.469 & 0.23 & 0.35& 0.560 \\
 $\eta_b(2S)$&  0.065 & 0.121 &  & 0.07 & 0.15& 0.269 \\
 $\eta_b(3S)$&  0.040 & 0.083 &   & 0.04 & 0.10& 0.208\\
 $\chi_{b0}(1P)$& 0.049 & 0.039 & 0.060  &  0.075& 0.038&\\
 $\chi_{b0}(2P)$& 0.038 & 0.029 &  & & 0.029&\\
 $\chi_{b0}(3P)$& 0.032 & 0.025 &  & && \\
\hline
\end{tabular}
\end{table}

\begin{figure}[!t]
\centering
 \includegraphics[width=0.48\textwidth]{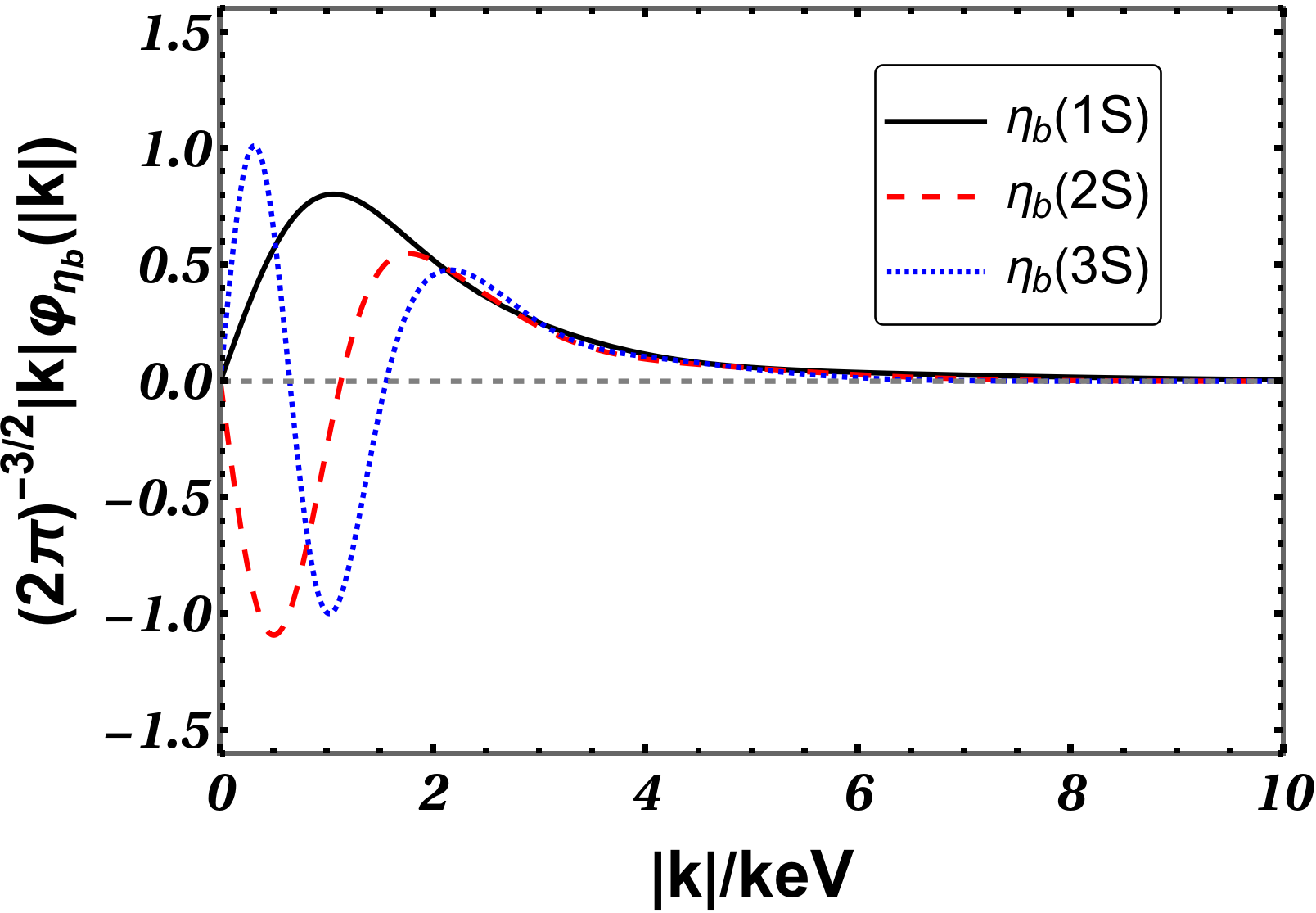}
 \includegraphics[width=0.48\textwidth]{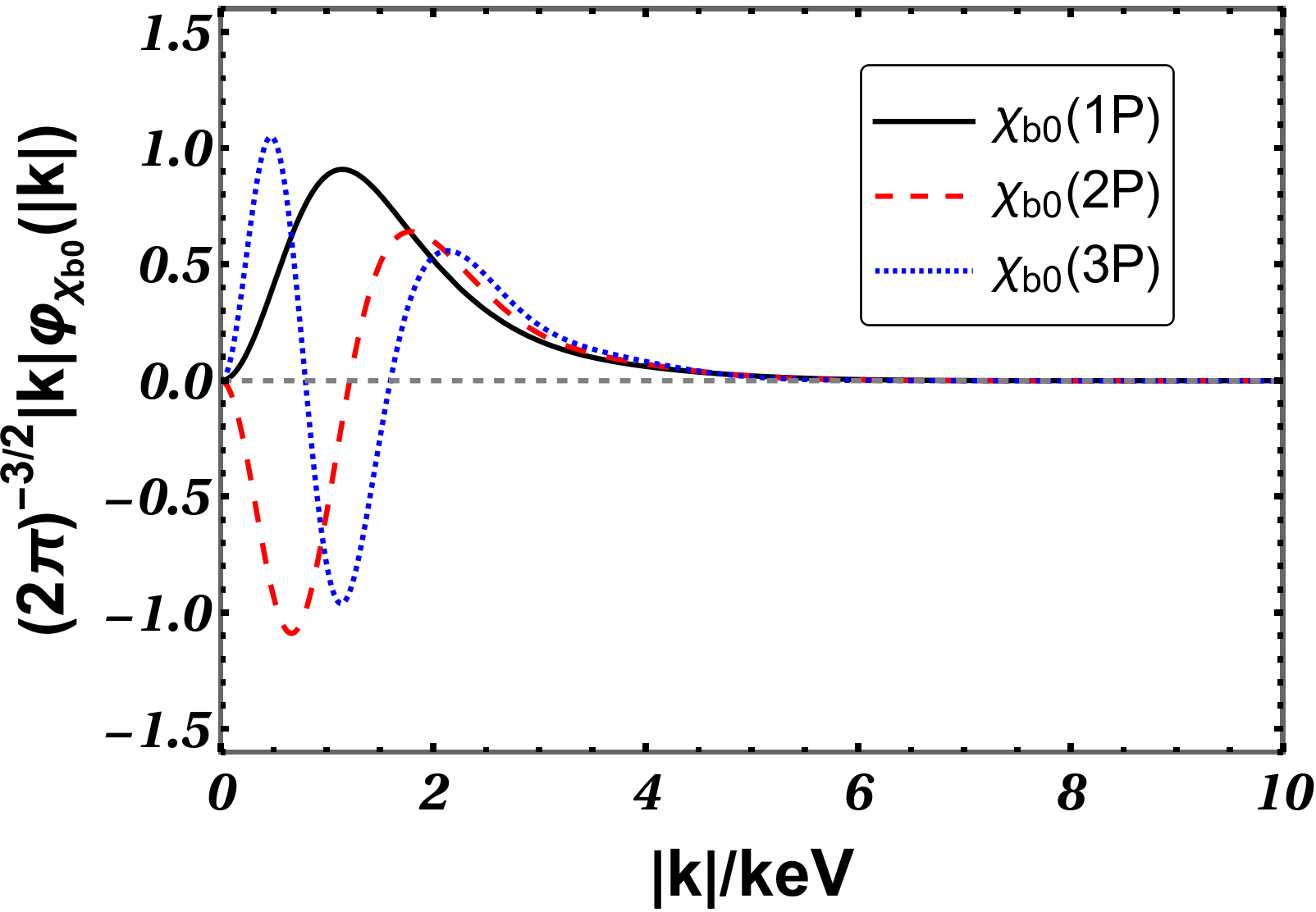}
 \caption{\label{fig:wfbb} (colored online) The radial wave function of $\eta_b(nS)$ (upper) and $\chi_{b0}(nP)$ (lower).}
\end{figure}

In this section we display and comment on the results of bottomonium, parallel to the charmonium.

The bottomonium spectrum is listed in Table \ref{tab:mbb}. The interaction parameters are fitted by the masses of $\eta_b(1S)$, $\varUpsilon(1S)$, $\varUpsilon(2S)$ and $\chi_{b0}(1P)$, so these four masses are inputs of our model, and all the other masses and the two-photon transition form factors are predictions. All the mesons in Table \ref{tab:mbb} are under the $B\bar{B}$ threshold, and the predicted masses agree with the observed values quite well, with an average error $12\text{ MeV}$. The radial wavefunctions of $\eta_b(nS)$ and $\chi_{b0}(nP)$ are in Fig. \ref{fig:wfbb}.

\subsection{The two-photon transition form factor and decay width}\label{subsec:formfactorBB}

\begin{figure}[!t]
\centering
 \includegraphics[width=0.48\textwidth]{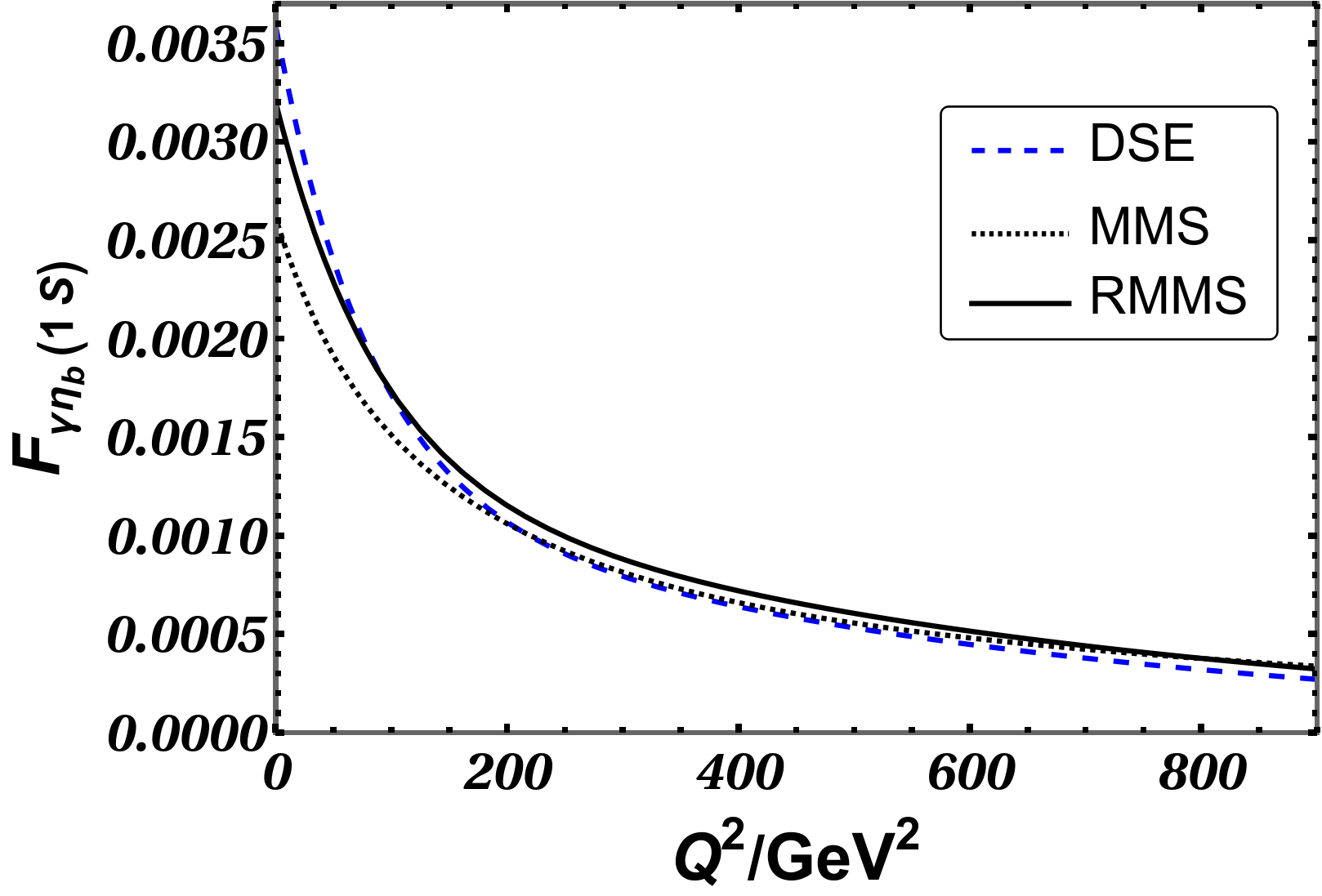}
 \includegraphics[width=0.48\textwidth]{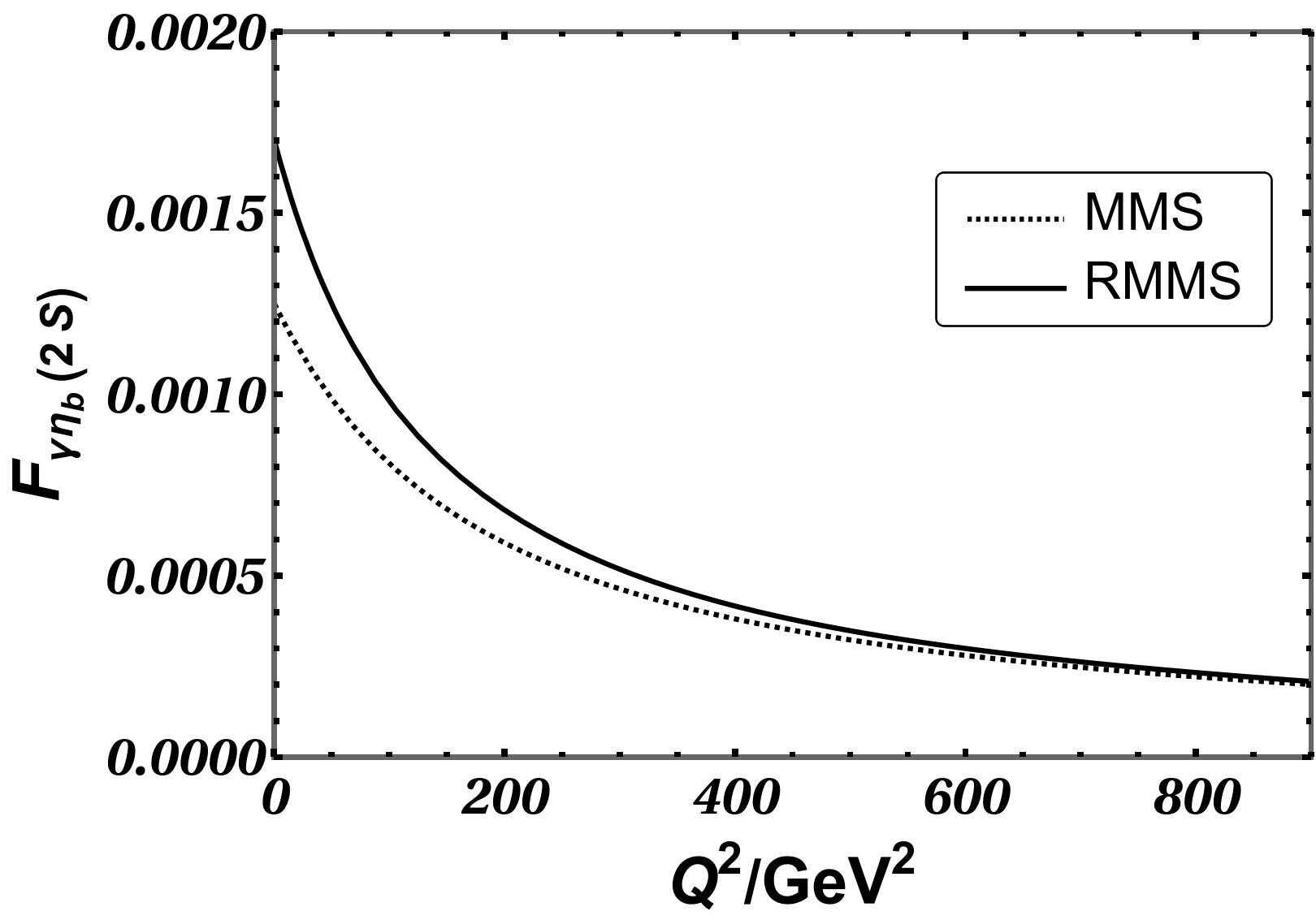}
 \includegraphics[width=0.48\textwidth]{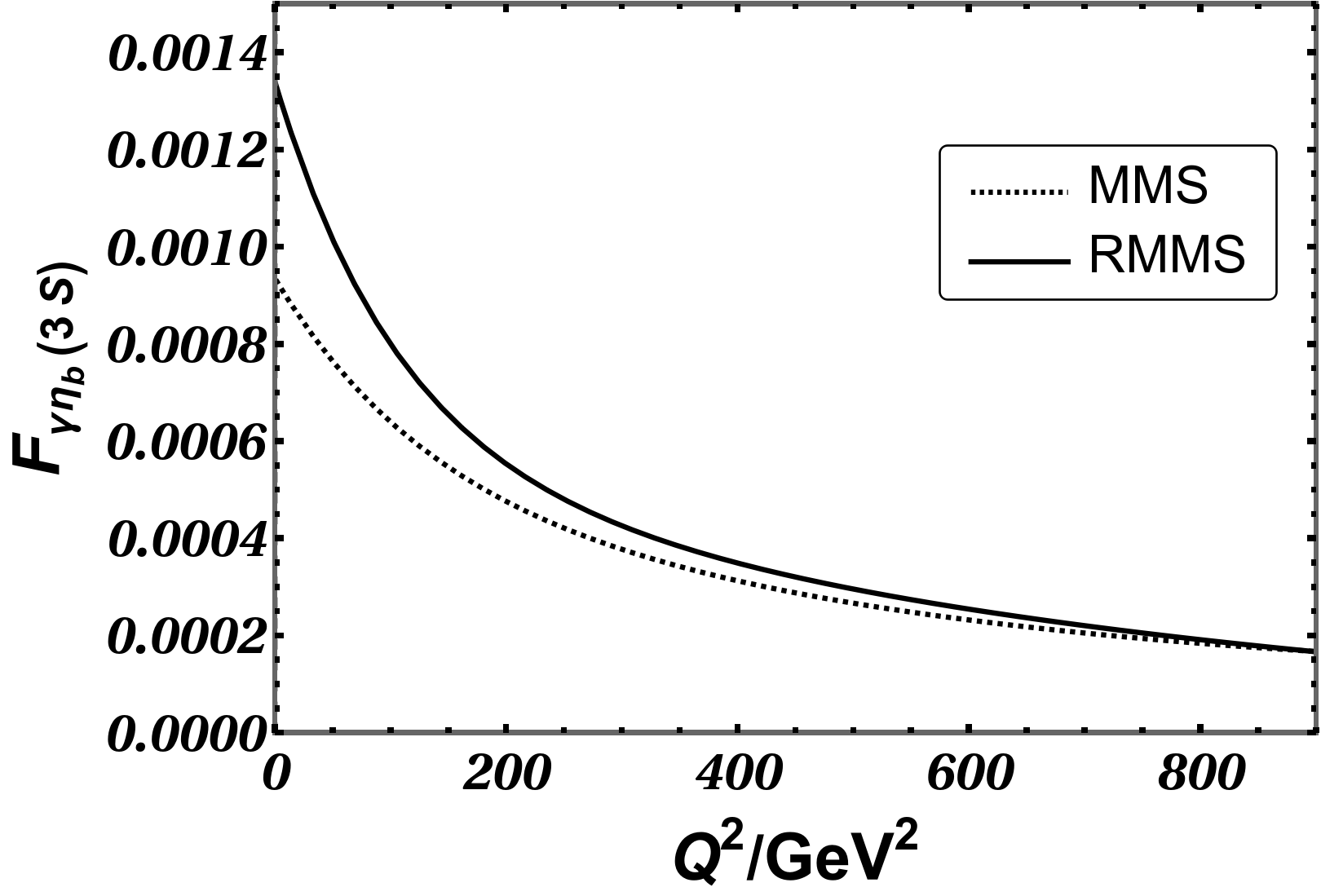}
 \caption{\label{fig:tffetab} (colored online) The two-photon transition form factor of $\eta_b(1S)$ (upper), $\eta_b(2S)$ (middle) and $\eta_b(3S)$ (lower). The black solid lines are our results with relativized mock meson state, Eq. (\ref{eq:RelativisedMockMeson}). The black dotted lines are our results with the old mock meson state, Eq. (\ref{eq:oldMockMeson}). The blue short-dashed line is the Dyson-Schwinger equation result \cite{Chen2017}.}
\end{figure}

\begin{figure}[!t]
\centering
 \includegraphics[width=0.48\textwidth]{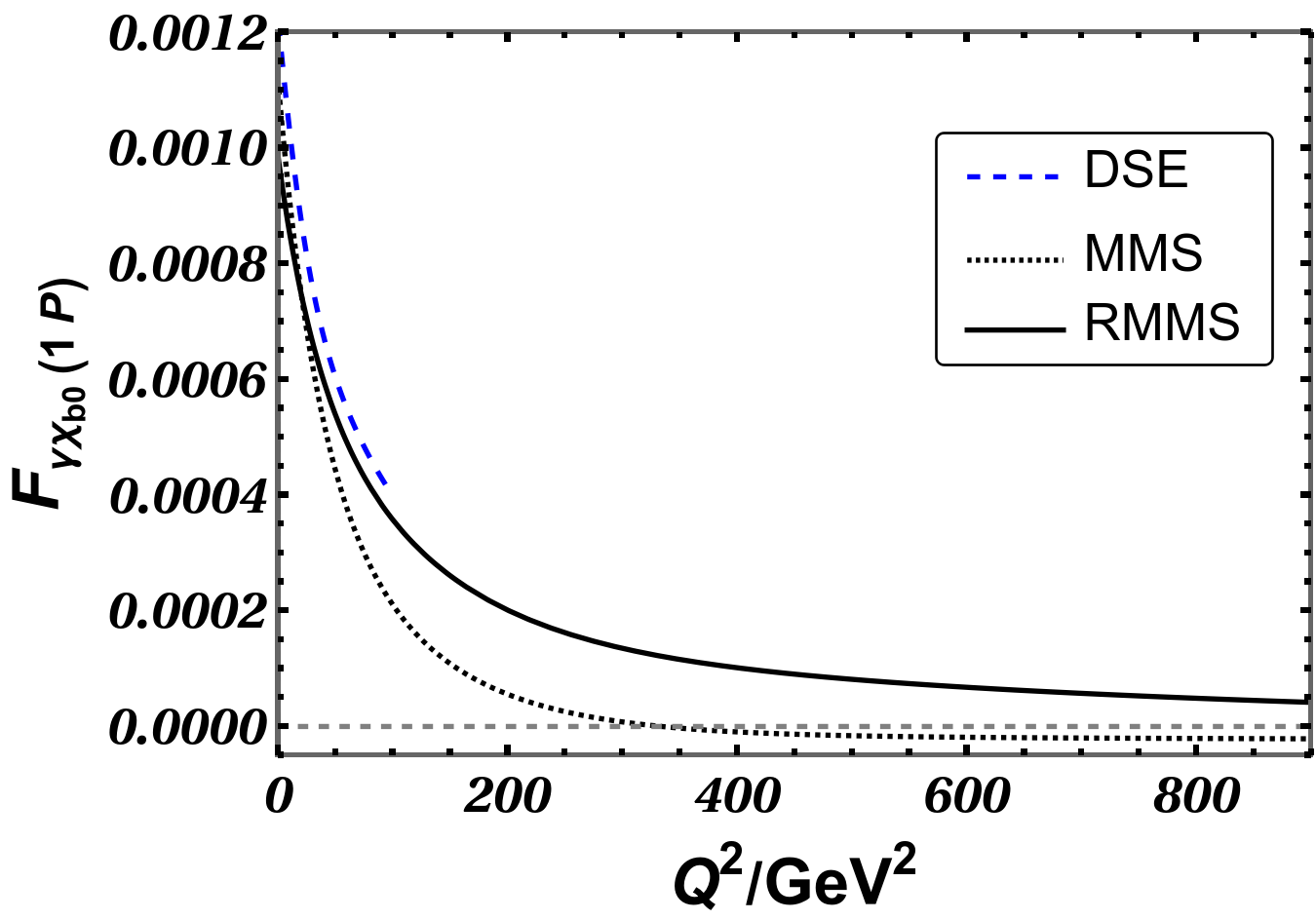}
 \includegraphics[width=0.48\textwidth]{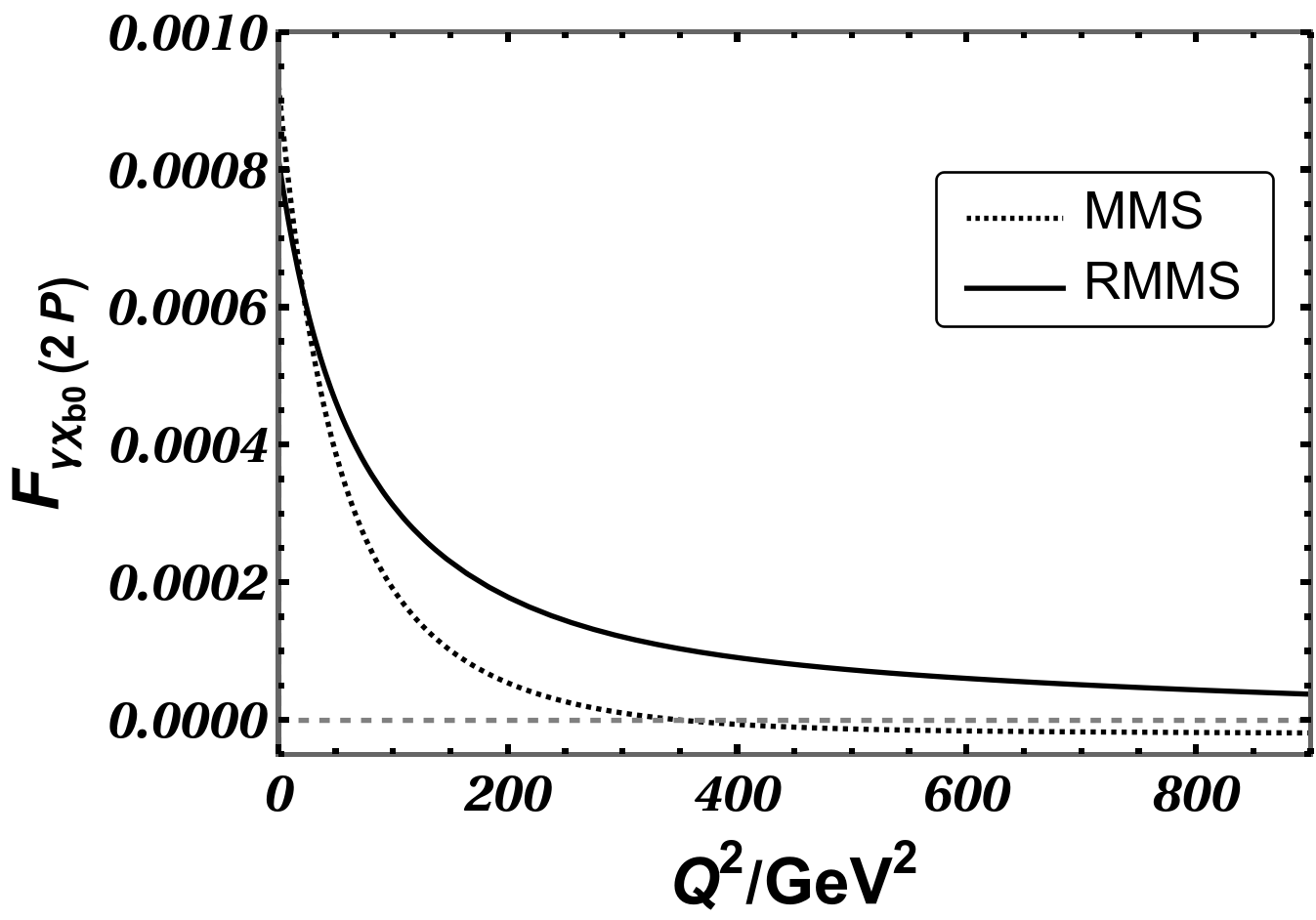}
 \includegraphics[width=0.48\textwidth]{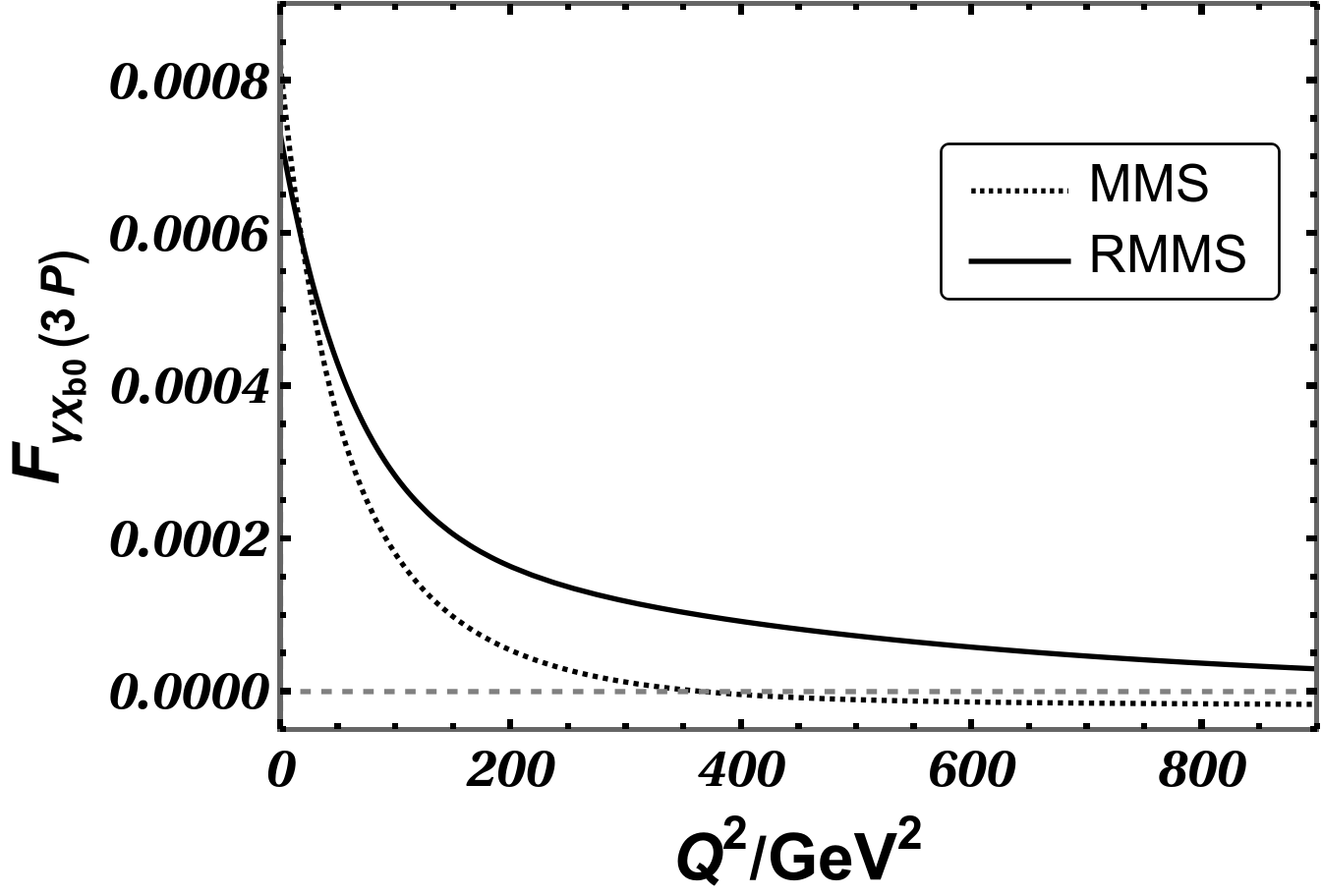}
 \caption{\label{fig:tffChib0} (colored online) The two-photon transition form factor of $\chi_{b0}(1P)$ (upper), $\chi_{b0}(2P)$ (middle) and $\chi_{b0}(3P)$ (lower). The black solid lines are our results with relativized mock meson state, Eq. (\ref{eq:RelativisedMockMeson}). The black dotted lines are our results with the old mock meson state, Eq. (\ref{eq:oldMockMeson}). The blue short-dashed line is the Dyson-Schwinger equation result \cite{Chen2017}.}
\end{figure}

The prediction of two-photon transition form factors of $\eta_b(nS)$ are in Fig. \ref{fig:tffetab} and those of $\chi_{b0}(nP)$ are in Fig. \ref{fig:tffChib0}. There is no experimental value for these quantities. The DSE results, which are proved to be credible, are available for $\eta_b(1S)$ and $\chi_{b0}(1P)$. In both cases, RMMS results are consistent with the DSE result, and MMS results deviate from the RMMS results.

Our predictions of two photon decay width of $F_{\gamma\eta_b}(nS)$ and $F_{\gamma\chi_{b0}}(nP)$
are listed in Table \ref{tab:DecayWidthbb}. There is no experimental value for these quantities presently. The RMMS results are more consistent with the DSE result, $\Gamma^{\textmd{DSE}}_{\gamma\gamma}$\\ \cite{Chen2017}, and relativistic quark model result, $\Gamma^{\textmd{RQM}}_{\gamma\gamma}$ \cite{Ebert2003}. The results from combination of non-relativistic quark model and an effective Lagrange, $\Gamma^{\textmd{NRQM}}_{\gamma\gamma}$ \cite{Lakhina2006}, and those from heavy-quark spin-symmetry approach, $\Gamma^{\textmd{HQSS}}_{\gamma\gamma}$ \cite{Lansberg2007}, are also listed to compare with.

\section{Summary and conclusion}\label{sec:conclusion}

In this work, we construct a relativized mock meson state, where Dirac spinors import kinetic relativistic correction and the dynamic wave function is the same as that solved from Schrödinger equation. Two-photon transition form factors of $\eta_{c,b}(n^1S_0)$ and $\chi_{c,b}(n^3P_0)$ ($n=1,2,3$) are calculated with both the old mock meson state and the relativized mock meson state. We compare our results with the experiment results, Dyson-Schwinger equation results, the basis light front quantization results, and some other theoretical prediction. It turns out that the kinetic relativistic correction imported by Dirac spinors is crucial to study the two-photon transion form factor in constituent quark model. The RMMS results provide a credible prediction for the two-photon transition form factors and decay widths of $\eta_{c,b}(n^1S_0)$ and $\chi_{c,b}(n^3P_0)$ ($n=1,2,3$).

In the next step, the effects of the relativized mock meson state on single quark form factors, two-photon transition form factors of higher spin mesons and the hadron transition form factors are to be studied.

\section*{Acknowledgments}
We acknowledge helpful conversations with Xian-Hui Zhong. This work is supported by: the Science Foundation of education department of Hunan province, China under contracts No. 24B0067.

\setcounter{section}{0}
\renewcommand{\thesection}{Appendix \Alph{section}}
\renewcommand{\thesubsection}{\Alph{section}\arabic{subsection}}

\section{non-relativistic limit of the relativized mock meson state}\label{sec:appendixA}

\setcounter{equation}{0}
\renewcommand{\theequation}{A\arabic{equation}}
\setcounter{figure}{0}
\setcounter{table}{0}
\renewcommand{\thefigure}{A\arabic{figure}}
\renewcommand{\thetable}{A\arabic{table}}

\subsection{The pseudoscalar meson}

The pseudoscalar relativized mock meson state is
\begin{eqnarray}\nonumber
|P(p) \rangle &=& \sqrt{\frac{2E_p}{N_c}} \sum_{\bm{s,\bar{s}}} \int\frac{d^3\bm{k} d^3\bm{\bar{k}}}{(2\pi)^6}\frac{1}{\sqrt{2E_{\bm{k}}}} \frac{1}{\sqrt{2E_{\bar{\bm{k}}}}}\\\label{eq:RMMS1S0}
&&\phi_P(k_r,p) \bar{u}(\bm{k},\bm{s}) \gamma_5 v(\bar{\bm{k}},\bar{\bm{s}}) b^\dag_{\bm{ks}} d^\dag_{\bm{\bar{k}\bar{s}}} | 0 \rangle.
\end{eqnarray}
In the Dirac representation, the spinors are
\begin{eqnarray}
u(\bm{k},\bm{s}) &=& \sqrt{E_{\bm{k}} + m}
\begin{pmatrix}
 1 \\ \frac{\bm{\sigma}\cdot \bm{k}}{E_{\bm{k}} + m}
\end{pmatrix}
 \xi^s, \\
 v(\bar{\bm{k}},\bar{\bm{s}}) &=& \sqrt{\bar{E}_{\bar{\bm{k}}} + \bar{m}}
\begin{pmatrix}
\frac{\bm{\sigma}\cdot \bar{\bm{k}}}{\bar{E}_{\bar{\bm{k}}} + \bar{m}} \\ 1
\end{pmatrix}
 \eta^{\bar{s}},
\end{eqnarray}
where $\bm{\sigma}$ is the Pauli matrix, $\xi^s$ and $\eta^{\bar{s}}$ are two-compoment spinors for quark and antiquark. In the non-relativistic limit, $\bm{k} \to 0$ and $\bar{\bm{k}} \to 0$, Eq. (\ref{eq:RMMS1S0}) reduces to
\begin{equation}\label{eq:ReducedRMMS1S0}
|P(p) \rangle = \sqrt{\frac{2E_p}{N_c}} \int\frac{d^3\bm{k} d^3\bm{\bar{k}}}{(2\pi)^6} \phi_P(k_r,p) \sqrt{2}\chi^{00}_{\bm{s\bar{s}}} b^\dag_{\bm{ks}} d^\dag_{\bm{\bar{k}\bar{s}}} | 0 \rangle,
\end{equation}
where $\chi^{00}_{\bm{s\bar{s}}} = \frac{1}{\sqrt{2}} (\xi^\uparrow\eta^\downarrow - \xi^\downarrow\eta^\uparrow)$ is the singlet spin wave function. Comparing Eq. (\ref{eq:ReducedRMMS1S0}) with Eq. (\ref{eq:oldMockMeson}), we get
\begin{equation}
 \phi_P(k_r,p)= \frac{\varphi_P\left(\bm{k}_r\right)}{\sqrt{2}} (2\pi)^3\delta^{(3)}(\bm{k}+\bm{\bar{k}}-\bm{p}).
\end{equation}

\subsection{The scalar meson}

The scalar relativized mock meson state is
\begin{eqnarray}\nonumber
|S(p) \rangle &=& \sqrt{\frac{2E_p}{N_c}} \sum_{\bm{s,\bar{s}}} \int\frac{d^3\bm{k} d^3\bm{\bar{k}}}{(2\pi)^6}\frac{1}{\sqrt{2E_{\bm{k}}}} \frac{1}{\sqrt{2E_{\bar{\bm{k}}}}}\\\label{eq:RMMS3P0}
&&\phi_S(k_r,p) \bar{u}(\bm{k},\bm{s})v(\bar{\bm{k}},\bar{\bm{s}}) b^\dag_{\bm{ks}} d^\dag_{\bm{\bar{k}\bar{s}}} | 0 \rangle.
\end{eqnarray}
In the non-relativistic limit, Eq. (\ref{eq:RMMS3P0}) reduces to
\begin{eqnarray}\nonumber
|S(p) \rangle &=& \sqrt{\frac{2E_p}{N_c}} \sum_{\bm{s,\bar{s}}} \int\frac{d^3\bm{k} d^3\bm{\bar{k}}}{(2\pi)^6} \frac{\sqrt{8\pi}}{2\tilde{m}}|\bm{k}_r|\\\label{eq:ReducedRMMS3P0}
&&\phi_S(k_r,p)  \chi^{1m_z}_{\bm{s\bar{s}}}Y^{1-m_z} b^\dag_{\bm{ks}} d^\dag_{\bm{\bar{k}\bar{s}}} | 0 \rangle,
\end{eqnarray}
where $\tilde{m} = \frac{m\bar{m}}{m+\bar{m}}$ is the reduced quark mass, $\chi^{1m_z}_{\bm{s\bar{s}}}Y^{1-m_z} = \frac{1}{\sqrt{3}} (\xi^\uparrow \eta^\uparrow Y^{1-1} - \frac{\xi^\uparrow\eta^\downarrow - \xi^\downarrow\eta^\uparrow}{\sqrt{2}} Y^{10} +\xi^\downarrow \eta^\downarrow Y^{1+1})$ is just the triplet spin wave function and the first-order spherical harmonic function combined by the Clebsch-Gordan coefficient. Comparing Eq. (\ref{eq:ReducedRMMS3P0}) with Eq. (\ref{eq:oldMockMeson}), we get
\begin{equation}\label{eq:NRL-RMMS3P0}
 \phi_S(k_r,p)= \frac{2\tilde{m}\varphi_S\left(|\bm{k}_r|\right)}{\sqrt{8\pi}|\bm{k}_r|} (2\pi)^3\delta^{(3)}(\bm{k}+\bm{\bar{k}}-\bm{p}).
\end{equation}
In order to make sure the single quark elastic form factor being normalized correctly, $\tilde{m}$ is replaced by $\tilde{E} = \frac{E_{\bm{k}} \bar{E}_{\bar{\bm{k}}}}{E_{\bm{k}} + \bar{E}_{\bar{\bm{k}}}}$.

\section{Two-photon transition form factors}\label{sec:appendixB}

\setcounter{equation}{0}
\renewcommand{\theequation}{B\arabic{equation}}
\setcounter{figure}{0}
\setcounter{table}{0}
\renewcommand{\thefigure}{B\arabic{figure}}
\renewcommand{\thetable}{B\arabic{table}}

We display the expressions of the two-photon transition form factor in the meson static frame, $p = (M, \vec{0})$. The 4-momentum of the two photons are
\begin{eqnarray}
 q_1 = (q_1^t,0,0,q_1^z) &=& (\frac{M^2+Q_1^2 - Q_2^2}{2M},0,0,\lambda),\\
 q_2 = (q_2^t,0,0,q_2^z) &=& (\frac{M^2-Q_1^2 + Q_2^2}{2M},0,0,-\lambda),\end{eqnarray}
where $Q_1^2 = q_1^2$, $Q_2^2 = q_2^2$ are the 4-momentum square and $\lambda = \frac{1}{2M}\{ M^4+Q_1^4+Q_2^4 -2(M^2Q_1^2 + M^2Q_2^2 + Q_1^2Q_2^2) \}^{1/2}$.

\subsection{$F_{\gamma\eta_c}$ from old mock meson state}

The two-photon symmetry form of the form factor is
\begin{eqnarray}\nonumber
&& \int d^4 x d^4 y e^{iq_1\cdot x+iq_2\cdot y}\langle 0|T\{ j^\mu(x) j^\nu(y)\}|M(p)\rangle \\\label{eq:symTFF1S0}
& =&\!\! \epsilon^{\mu\nu\alpha\beta} q_{1\alpha}q_{2\beta}F_{P\gamma\gamma}(q_1^2,q_2^2)(2\pi)^4\delta^{(4)}(q_1\!+\!q_2\!-\!p).
\end{eqnarray}
The left hand side
\begin{eqnarray}\nonumber
 \text{LHS}&=&\int d^4 x d^4 y e^{iq_1\cdot x+iq_2\cdot y}\langle 0| j^\mu(x) j^\nu(y)|M(p)\rangle|_{x^0>y^0}\\\nonumber
 &&+\{x \leftrightarrow y\}\\\nonumber
 &=&\int d^4 x d^4 y e^{iq_1\cdot x}e^{i(q_2\cdot y-p\cdot y+q_1\cdot y)}\\\nonumber
 &&\cdot\langle 0| j^\mu(x) j^\nu(0)|M(p)\rangle|_{x^0>0} +\{q_1,\mu \leftrightarrow q_2,\nu\}\\\nonumber
 &=&(2\pi)^4\delta^{(4)}(q_1+q_2-p)\int d^4 x e^{iq_1\cdot x}\\\nonumber
 &&\cdot\langle 0| j^\mu(x) j^\nu(0)|M(p)\rangle|_{x^0>0} +\{q_1,\mu \leftrightarrow q_2,\nu\}.
\end{eqnarray}
So that, we get the symmetric form
\begin{eqnarray}\nonumber
&&\int d^4 x e^{iq_1\cdot x}\langle 0| j^\mu(x) j^\nu(0)|M(p)\rangle|_{x^0>0} +\{q_1,\mu \leftrightarrow q_2,\nu\}\\
&&= M^{\mu\nu}=\epsilon^{\mu\nu\alpha\beta} q_{1\alpha}q_{2\beta}F_{P\gamma\gamma}(q_1^2,q_2^2).
\end{eqnarray}
It's worth mentioning that the time integration is operated as, e.g., $\int_0^{\infty} dx^0 e^{i(q^t_1-E_{\bm{k}}-E_{\bm{k}-\bm{q}_1}) x^0} =\int_0^{\infty(1-i\epsilon)} dx^0 e^{i(q^t_1-E_{\bm{k}}-E_{\bm{k}-\bm{q}_1}) x^0} = \frac{i}{q^t_1-E_{\bm{k}}-E_{\bm{k}-\bm{q}_1}}$. Let $\{\mu,\nu\} = \{1,2\}$ or $\{2,1\}$, after some manipulation of the Dirac matrices, we get the expression of two-phonton transion form factor
\begin{eqnarray}\nonumber
 F_{P\gamma\gamma}(q_1^2,q_2^2) =\frac{e_q^2 \sqrt{2MN_c}}{q_1^t q_2^z - q_2^t q_1^z} \int \frac{d^3\bm{k}}{(2\pi)^3}\frac{\varphi_P(\bm{k})}{E_{\bm{k}} E_{\bm{k}+\bm{q_1}}}\times \\
 (\frac{1}{E_{\bm{k}} \!+\! E_{\bm{k}+\bm{q_1}} \!-\! q_1^t} \!+\! \frac{1}{E_{\bm{k}} \!+\! E_{\bm{k}+\bm{q_1}} \!-\! q_2^t})\sqrt{2}m(2\bm{k}^z \!+\! q_1^z).
\end{eqnarray}
The decay width is
\begin{equation}
 \Gamma_{P\gamma\gamma} = \frac{\pi \alpha^2 M^3}{4}|F_{P\gamma\gamma}(0,0)|^2.
\end{equation}

\subsection{$F_{\gamma\eta_c}$ from relativized mock meson state}

\begin{eqnarray}\nonumber
 F_{P\gamma\gamma}(q_1^2,q_2^2) &=&\frac{e_q^2 \sqrt{2MN_c}}{q_1^t q_2^z - q_2^t q_1^z}
 \left\{ \int \frac{d^3\bm{k}}{(2\pi)^3}\frac{\varphi_P(\bm{k})}{\sqrt{2}}[A_1 + A_2]\right. \\
&& \left. -\{q_1 \leftrightarrow q_2\}
 \right\},
\end{eqnarray}
where $A_1 = \frac{(m-\bar{m})E_{\bm{k} + \bm{q}_1} \bm{k}^z - (m-\bar{m})E_{\bm{k}} \bm{k}^z + m E_{\bm{\bar{k}}} \bm{q}_1^z  + \bar{m} E_{\bm{k}} \bm{q}_1^z}{2E_{\bm{k}} \bar{E}_{\bar{\bm{k}}} E_{\bm{k} + \bm{q}_1} (\bar{E}_{\bar{\bm{k}}} + E_{\bm{k} + \bm{q}_1} -q_1^t)} $, $A_2 = \frac{(m-\bar{m})\bar{E}_{\bm{k} - \bm{q}_1} \bm{k}^z - (m-\bar{m})\bar{E}_{\bm{k}} \bm{k}^z + m E_{\bm{\bar{k}}} \bm{q}_1^z  + \bar{m} E_{\bm{k}} \bm{q}_1^z}{2E_{\bm{k}} \bar{E}_{\bar{\bm{k}}} \bar{E}_{\bm{k} - \bm{q}_1} (E_{\bm{k}} + \bar{E}_{\bm{k} - \bm{q}_1} -q_1^t)} $.

\subsection{$F_{\gamma\chi_{c0}}$ from old mock meson state}

The two-photon transition form factor of scalar meson is defined as \cite{Hoferichter2020}
\begin{eqnarray}\nonumber
 M^{\mu\nu} &=& \int d^4 x e^{iq_1\cdot x}\langle 0|T\{ j^\mu(x) j^\nu(0)\}|M\rangle \\
&=&  T^{\mu\nu}_1 F_1(q_1^2,q_2^2) + T^{\mu\nu}_2 F_2(q_1^2,q_2^2),
\end{eqnarray}
where
\begin{eqnarray}
 T^{\mu\nu}_1 &=& q_1\!\cdot\! q_2 g^{\mu\nu} - q_2^\mu q_1^\nu,\\
 T^{\mu\nu}_2 &=& \!q_1^2 q_2^2 g^{\mu\nu} \!+\! q_1\!\cdot\! q_2 q_1^\mu q_2^\nu \!-\! q_1^2 q_2^\mu q_2^\nu \!-\! q_2^2 q_1^\mu q_1^\nu,
\end{eqnarray}
with $g^{\mu\nu} = \text{diag} \{1,-1,-1,-1\}$. In the single-tag case, $q_1^2 = -Q^2$, $q_2^2 = 0$, $F_{\gamma S}(Q^2) = F_1(-Q^2,0)$ could be calculated with $\{\mu,\nu \} = \{1,1 \}$ or $\{2,2\}$.
\begin{equation}
 F_{\gamma S}(Q^2) = \frac{e_q^2 2\sqrt{2M N_c}}{M^2+Q^2} \!\int\! \frac{d^3\bm{k}}{(2\pi)^3} \frac{-i \varphi_S(|\bm{k}|) \cdot A_3}{4E_{\bm{k}} E_{\bm{k}+\bm{q_1}} (E_{\bm{k}} \!+\! E_{\bm{k}+\bm{q_1}} \!-\! q_1^t)},
\end{equation}
where $A_3 = -\frac{2\sqrt{2}|\bm{k}|m}{\sqrt{\pi}} -\frac{\sqrt{2}\bm{k}^z/|\bm{k}|}{\sqrt{\pi}(E_{\bm{k}} + m)}[(E_{\bm{k}} + m)^2(\bm{q}_1^z + \bm{k}^z) + \bm{q}_1^z\bm{k}^2 -2\bm{q}_1^z(\bm{k}^z)^2 + \bm{k}^2\bm{k}^z -2(\bm{k}^z)^3] + \frac{2\sqrt{2}(1-(\bm{k}^z/|\bm{k}|)^2)}{\sqrt{\pi}(E_{\bm{k}} + m)}|\bm{k}|(\bm{k}\cdot\bm{q}_1) (\bm{k}^z)^2$.
The decay width is
\begin{equation}
 \Gamma_{S\gamma\gamma} = \frac{\pi \alpha^2 M^3}{4}|F_{\gamma S}(0)|^2.
\end{equation}

\subsection{$F_{\gamma\chi_{c0}}$ from relativized mock meson state}

\begin{eqnarray}\nonumber
 F_{\gamma S}(Q^2) &=& \frac{e_q^2 2\sqrt{2M N_c}}{M^2+Q^2} \!\int\! \frac{d^3\bm{k}}{(2\pi)^3} \frac{2E_{\bm{k}} \bar{E}_{\bar{\bm{k}}}}{E_{\bm{k}} + \bar{E}_{\bar{\bm{k}}}}\frac{\varphi_S\left(|\bm{k}_r|\right)}{\sqrt{8\pi}|\bm{k}_r|}\\
 && \times
  (-i)[A_4 + A_5] + \{ q_1 \leftrightarrow q_2 \},
\end{eqnarray}
where $A_4 = [2E_{\bm{k}} \bar{E}_{\bar{\bm{k}}} E_{\bm{k} + \bm{q}_1} (\bar{E}_{\bar{\bm{k}}} + E_{\bm{k} + \bm{q}_1} -q_1^t)]^{-1} \times \{ m^2\bar{m} + m[E_{\bm{k} + \bm{q}_1} \bar{E}_{\bm{k}} - E_{\bm{k}} \bar{E}_{\bar{\bm{k}}} + \bm{q}_1\cdot\bm{k} -2(\bm{k}^x)^2] - \bar{m} [E_{\bm{k} + \bm{q}_1}E_{\bm{k}} -(\bm{k}+\bm{q}_1)\cdot\bm{k} + 2(\bm{k}^x)^2] \}$, $A_5 = [2E_{\bm{k}} \bar{E}_{\bar{\bm{k}}} \bar{E}_{\bm{k} - \bm{q}_1} (E_{\bm{k}} + \bar{E}_{\bm{k} - \bm{q}_1} -q_1^t)]^{-1} \times \{ m\bar{m}^2 - m[\bar{E}_{\bm{k} - \bm{q}_1} \bar{E}_{\bar{\bm{k}}} + (\bm{q}_1-\bm{k})\cdot\bm{k} +2(\bm{k}^x)^2 ] + \bar{m}[\bar{E}_{\bm{k} - \bm{q}_1} E_{\bm{k}} - E_{\bm{k}} \bar{E}_{\bar{\bm{k}}} -\bm{q}_1\cdot\bm{k} -2(\bm{k}^x)^2] \}$.

\bibliographystyle{unsrt}
\bibliography{../references/TFFcc}

\end{document}